\documentclass[%
 reprint, superscriptaddress, amsmath,amssymb, aps]{revtex4-2}

\usepackage{graphicx}
\usepackage{dcolumn}
\usepackage{bm}
\usepackage{amsmath} 
\usepackage{siunitx}
\usepackage{float}
\usepackage{physics} 
\usepackage{rotating} 
\usepackage{colortbl} 
\usepackage{booktabs} 
\usepackage{multirow} 
\usepackage{diagbox} 
\usepackage{makecell} 
\usepackage{xcolor} 
\usepackage{placeins}

\begin{document}

\preprint{APS/123-QED}

\title{Dielectric Loaded Waveguide Terahertz LINACs}

\author{Mostafa Vahdani}
\email{mostafa.vahdani@desy.de}
\affiliation{%
	Center for Free-Electron Laser Science CFEL,\\Deutsches Elektronen-Synchrotron DESY, Notkestrasse 85, 22607 Hamburg, Germany
}%
\affiliation{%
	Department of Physics, Universität Hamburg, Jungiusstrasse 9, 20355 Hamburg, Germany
}
\author{Moein Fakhari}%
 \affiliation{%
Center for Free-Electron Laser Science CFEL,\\Deutsches Elektronen-Synchrotron DESY, Notkestrasse 85, 22607 Hamburg, Germany
}%
\author{Franz X. Kärtner}
\affiliation{%
	Center for Free-Electron Laser Science CFEL,\\Deutsches Elektronen-Synchrotron DESY, Notkestrasse 85, 22607 Hamburg, Germany
}
\affiliation{%
	Department of Physics, Universität Hamburg, Jungiusstrasse 9, 20355 Hamburg, Germany
}
\affiliation{%
	The Hamburg Center for Ultrafast Imaging, Luruper Chaussee 149, 22761 Hamburg, Germany
}

\begin{abstract}
Dielectric loaded waveguides (DLWs) driven by multicycle terahertz (THz) pulses hold great promise as compact linear accelerators (LINACs) due to their ability to sustain higher breakdown fields at THz frequencies compared to conventional RF components. Precise control of the THz pulse's phase and group velocities within the DLW can be achieved by adjusting the dimensions of the dielectric tube. Optimization of the DLW parameters has to take various factors into account like initial electron energy, THz pulse energy available, THz pulse width, DLW dimensions, etc. To maximize the final kinetic energy of the accelerated electrons, this study presents a comprehensive analytical/numerical guideline for cylindrical DLW LINACs. Additionally, graphic representations are introduced to visualize optimal designs for varying initial electron and THz pulse energies. The provided guideline figures enable designers to tailor the accelerator to specific requirements, paving the way for potential advancements in THz-driven particle acceleration and offering cost-effective alternatives to conventional RF accelerators for various scientific, industrial, and medical applications. 
\end{abstract}

\maketitle

\section{\label{sec:Introduction}Introduction}

Conventional Radio Frequency (RF) accelerators operate by using alternating electric fields to accelerate charged particles, such as electrons, to high energies. A higher acceleration gradient is always preferable and will result in an increase in bunch quality since space charge affects the beam, especially as long as the beam is sub-relativistic. In order to increase the acceleration gradient, one needs to increase the accelerating electric field which is limited by field-emission at metal surfaces (breakdown voltage) in conventional RF accelerators. It was always a dream for accelerator physicists to go to higher operating frequencies because of the expected higher break down fields \cite{dal2016rf,dal2016experimental,thompson2008breakdown}. Laser driven accelerators such as Dielectric Laser Accelerators (DLA) can achieve GV/m accelerating fields. However, the short wavelength of the fields and small cross sectional area of the accelerator significantly limits the bunch charge \cite{peralta2013demonstration,breuer2013laser,carbajo2016direct}.  Extremely high timing precision is also required for injecting electron bunches due to short wavelength of optical fields. The short wavelength also results in an increased energy spread and emittance growth. Other methods such as plasma-based acceleration were developed to overcome these limitations \cite{malka2008principles,leemans2009laser,guenot2017relativistic,salehi2017mev,he2013high}. Plasma waves can be produced by intense laser pulses or high charge electron bunches in plasmas. Laser Plasma Accelerators (LPA) suffer however from plasma instabilities and as recently shown need exquisite control on the drive laser pulses to achieve high energy stability and good beam quality \cite{kirchen2021optimal,jalas2021bayesian,ferran2022energy}. This is where Terahertz (THz) frequencies come into play, offering a potential solution to mitigate these limitations, since the accelerating structure is still a solid state device \cite{nanni2015terahertz,walsh2017demonstration,curry2017relativistic,zhang2018segmented,wong2017laser}. Another advantage of THz frequencies lies in their ability to allow for higher field gradients due to the shorter pulse duration available at higher frequencies and the shorter wavelength when compared to RF-waves. Still significant charge can be accelerated with excellent beam quality, stability and most importantly lower emittance or higher brightness due to the increased field strength, making them an attractive candidate for particle accelerators. Break down fields at THz frequencies, e.g. 100-300 GHz can be considerably higher when compared to the RF at 3 GHz using the scaling rules empirically confirmed \cite{loew1988rf,kilpatrick1957criterion,dobert2005high}. Also, THz sources both in the single-cycle and multi-cycle regime are becoming increasingly mature \cite{wu2023generation}, but are yet by far not as high developed as RF-sources.

The advancements in THz-driven accelerators can potentially lead to more compact and cost-effective acceleration systems, offering an alternative to conventional RF accelerators and enabling new scientific discoveries, industrial applications, and medical treatments that rely on high-energy particle beams. In recent years THz-driven accelerators have emerged as a promising field of research, driven by advancements in THz source technologies. Many groups are actively exploring various methods and techniques to harness the potential of THz-driven accelerators and beam manipulation devices.

THz accelerators driven by single-cycle THz pulses generated through nonlinear crystals have been studied and developed over the past decade \cite{zhang2018segmented,zhang2019femtosecond}. While multi-cycle THz accelerators employing resonant cavities have shown promise for accelerator device implementation \cite{othman2020experimental}, challenges still persist in fabricating metallic structures with the required surface quality at THz frequencies. Traveling wave linear accelerators present another type of particle accelerator in the THz spectrum and have long been considered a viable solution. The idea of using traveling waves for electron acceleration has been already explored for RF accelerators \cite{fry1947travelling}. The phase velocity of the traveling wave in the accelerator waveguide must be reduced to match the electron bunch velocity for optimal interaction between electrons and the electromagnetic wave, facilitating efficient acceleration. This reduction is typically achieved through the utilization of slow wave structures like corrugated waveguides or Dielectric Loaded Waveguides (DLW). THz traveling wave can be generated either externally, for example, by nonlinear crystals or internally via a colinear intense driving beam using Cerenkov radiation (wake fields) \cite{ginsburg1947use,gai1988experimental,cook2009observation}. It is noted that in the configuration of two-beam accelerators (TBA), a modified version of colinear wakefield accelerators, the generated electromagnetic wave is transferred to a separate waveguide hosting the main beam (witness beam) \cite{gai2001experimental,jing2016dielectric}.

DLWs have emerged as a particularly attractive option for multicycle THz acceleration, primarily due to their ease of fabrication. While the concept of employing DLWs for RF accelerators dates back to 1947 \cite{flesher1951dielectric,frankel1947tm}, their practicality and favorability have become more pronounced in the THz regime. During the last decade, many studies have focused on implementing THz accelerators using DLWs in both rectangular \cite{hibberd2020acceleration} and cylindrical structures \cite{nanni2015terahertz,zhang2020cascaded,xu2021cascaded,tang2021stable}. Cylindrical DLWs, in particular, offer superior field uniformity compared to their rectangular counterparts (will be discussed in this paper). In addition, it has been shown, that DLWs in conjunction with metallic wire waveguides and surface waves enables the efficient implementation of THz accelerators \cite{yu2024hollow,yu2023megaelectronvolt}.

In this paper our focus is on optimizing the structure of cylindrical DLWs for THz acceleration. The design of cylindrical DLW LINACs involves incorporating a dielectric layer to reduce the phase velocity, as the metallic waveguides exhibit phase velocities exceeding the speed of light. Furthermore, the addition of a dielectric layer unavoidably reduces the group velocity, limiting the interaction length between the electrons and the accelerating field within the DLW. In this context, multi-cycle THz pulses are used for increasing the interaction length, as they provide more cycles with lower bandwidth, thereby reducing the dispersion effect within the waveguide. In the simulations conducted, a rectangular THz pulse along the DLW is considered. If the pulse is sufficiently long and narrow-band, it is possible to neglect the dispersion effect of the waveguide. However, the envelope velocity of the pulse within the waveguide is still calculated and taken into account to ensure more accurate modeling and analysis. In the first section of this paper, we explore the electromagnetic characteristics of the DLW. Subsequently, in the second section, we focus on optimizing the DLW dimensions and THz pulse properties to maximize energy gain. We also present the key parameters that need to be initialized, varied, or optimized. Finally, in the third section, we briefly discuss the effects of the DLW LINAC on axial parameters of an electron bunch.

\section{\label{sec:1.	Electromagnetic field computation} Electromagnetic field computation}

\begin{figure*}
	\includegraphics[width=1\textwidth]{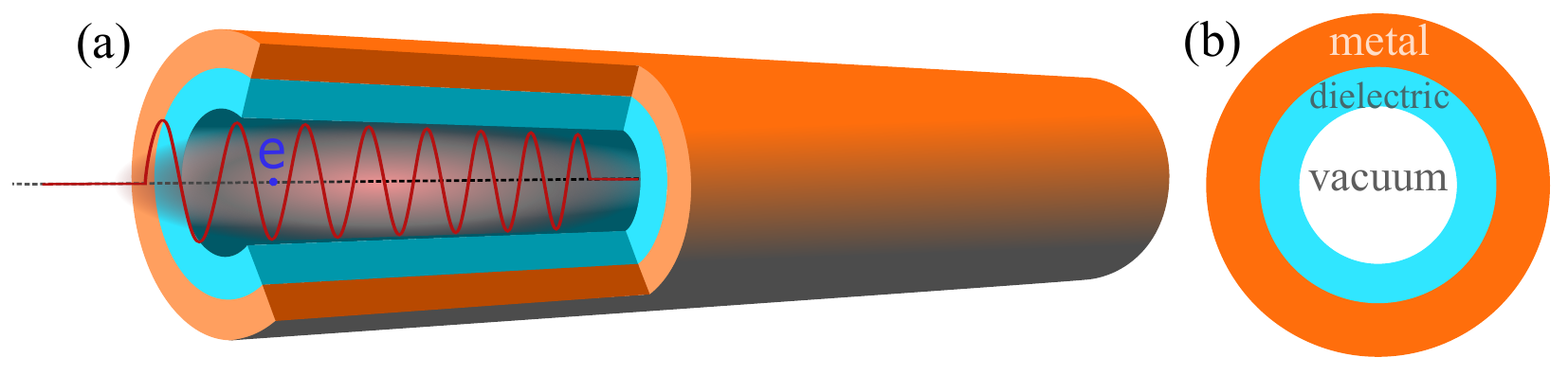}
	\caption{\label{fig:DLW LINAC/tube3}(a) DLW structure comprising a dielectric tube (light blue) surrounded by a metallic layer (orange) functioning as a THz LINAC (b) cross sectional view of the DLW.}
\end{figure*}

Here, we employ an analytical/numerical approach to investigate the electromagnetic fields in a cylindrical DLW LINAC which is depicted schematically in Fig.~\ref{fig:DLW LINAC/tube3}. The metallic layer has to be thick compared to the skin depth of the metal to effectively confine the fields within the structure.

For the DLW design and optimization of electron acceleration along the DLW, the exact solutions for the electromagnetic fields within the tube need to be found. Typically, the coupling between the electron beam current and the electromagnetic wave complicates the analysis. However, in the THz regime, where low charge single bunches are assumed, the influence of the beam current on the electromagnetic wave becomes negligible, allowing for decoupling of the two phenomena. Consequently, the problem becomes easier to solve using numerical and analytical methods. 

Solving Maxwell’s equations along with applying the boundary conditions and continuity of the electric and magnetic fields in the waveguide results in the waveguide modes. Field distributions with vanishing longitudinal electric field component are called Transverse Electric ($TE$) modes and those with vanishing longitudinal magnetic field are Transverse Magnetic ($TM$) modes. The DLW structure depicted in Fig.~\ref{fig:DLW LINAC/tube3} also supports Hybrid Electromagnetic Modes ($HEM$) due to having two media with different refractive indices. $HEM$ modes have both electric field and magnetic field longitudinal field components. 

The cylindrical ${TM}_{01}$ mode (we will explore the indices in the following section) is identified as the optimal mode for particle acceleration due to its strong longitudinal electric field along the axis for a given input power. The subsequent section presents a rigorous determination of the field distribution for this mode.   

\subsection{\label{sec:Field distribution}Field distribution}

From electromagnetic field theory we know that the electric and magnetic fields in a homogeneous source free region can be expressed in terms of the vector potential $\vec{A}$ as \cite{fields1961time}
\begin{subequations}
	\label{eq:Electric and magnetic field from potential}
	\begin{eqnarray}
		\vec{E}=-j\omega\mu\vec{A}+\frac{1}{j\omega\varepsilon}\mathrm{\nabla}\left(\mathrm{\nabla}\cdot\vec{A}\right)
	\end{eqnarray}
	\begin{equation}
		\vec{H}=\mathrm{\nabla}\times\vec{A}
	\end{equation}
\end{subequations}
There is an arbitrariness in the choice of vector potentials.  For $TM$ modes the magnetic vector potential $\vec{A}$ has only a longitudinal component, $\Psi$, i.e.
\begin{equation}
	\vec{A}={\vec{u}}_z\Psi
	\label{eq:potential and scalar potential}
\end{equation}
which obeys the scalar Helmholtz equation. Due to the cylindrical symmetry of the DLW structure, we write the Helmholtz equation for $\Psi$ in cylindrical coordinates.
\begin{equation}
\frac{1}{\rho}\frac{\partial}{\partial\rho}\left(\rho\frac{\partial\Psi}{\partial\rho}\right)+\frac{1}{\rho^2}\frac{\partial^2\Psi}{\partial\phi^2}+ \frac{\partial^2\Psi}{\partial z^2}+k^2\Psi=0,\   \ k^2=\omega^2\mu\varepsilon
\label{eq:Helmholtz}
\end{equation}
Here, $\epsilon$ and $\mu$ denote the medium's permittivity and permeability, respectively. By utilizing the method of separation of variables in cylindrical coordinates, the wave functions $\mathrm{\Psi}$ can be expressed as the product of functions $R\left(\rho\right), {\Phi}\left(\phi\right),$ and $Z\left(z\right)$.
\begin{equation}
\mathrm{\Psi}=R\left(\rho\right)\mathrm{\Phi}\left(\phi\right)Z\left(z\right)
\label{eq:Seperation of Variables}
\end{equation}
Substitution of $\mathrm{\Psi}$ into Eq.~(\ref{eq:Helmholtz}) and division by $\mathrm{\Psi}$ will result in
\begin{equation}
	\label{eq:Psi}
	\frac{1}{\rho R} \frac{d}{d\rho} \left( \rho \frac{dR}{d\rho} \right) + \frac{1}{\rho^2 \Phi} \frac{d^2 \Phi}{d\phi^2} + \frac{1}{Z} \frac{\partial^2 Z}{\partial z^2} + k^2 = 0
\end{equation}
Since the third term is independent of $\rho$ and $\phi$, and cannot be a function of $z$, it must be equal to a constant.
\begin{equation}
	\label{eq:Z}
	\frac{1}{Z} \frac{\partial^2 Z}{\partial z^2} = -k_z^2
\end{equation}
substituting Eq.~(\ref{eq:Z}) into Eq.~(\ref{eq:Psi}) and multiplying by $\rho^2$ yields:
\begin{equation}
	\label{eq:}
	\frac{\rho}{R} \frac{d}{d\rho} \left( \rho \frac{dR}{d\rho} \right) + \frac{1}{\Phi} \frac{d^2 \Phi}{d\phi^2} + (k^2 - k_z^2)\rho^2 = 0
\end{equation}	
The second term is solely a function of $\phi$ while the rest of terms are functions of $\rho$. Therefore 
\begin{equation}
	\label{eq:Phi}	
	\frac{1}{\Phi} \frac{d^2 \Phi}{d\phi^2} = -m^2
\end{equation}
The constant \( m \) must be an integer due to the periodic nature of \( \Phi(\phi) \) in cylindrical coordinates. Consequently, the equation for \( R(\rho) \) can be transformed into Bessel’s equation.
\begin{equation}
	\rho\frac{d}{d\rho}\left(\rho\frac{dR}{d\rho}\right)+\left((k^2-k_z^2)\rho^2-m^2\right)R=0.
	\label{eq:bessel}
\end{equation}

Eqs.~(\ref{eq:Z}) and (\ref{eq:Phi}) reveal that \( \Phi(\phi) \) and \( Z(z) \) are solutions of the harmonic equation, resulting in harmonic functions. \( R(\rho) \) in Eq.~(\ref{eq:bessel}) adheres to Bessel’s equation of order \( n \) and can be represented using Bessel functions of the first kind, \( J_m(k_\rho\rho) \), and Bessel functions of the second kind (Neumann function) \( Y_m(k_\rho\rho) \) or Hankel functions of the first and second kind \( H_m^{(1)}(k_\rho\rho) \) and \( H_m^{(2)}(k_\rho\rho) \), respectively. The term \( k^2-k_z^2 \) can be replaced by \( k_\rho^2 \) which corresponds to the radial component of the wave vector \( k \). Hence, the wave function \( \Psi \) in Eq.~(\ref{eq:Seperation of Variables}) is expressed in terms of the functions
\begin{equation}
	Z(z) \sim e^{-jk_zz}, e^{jk_zz}
	\label{eq:z_terms}
\end{equation}
\begin{equation}
	\Phi(\phi) \sim e^{-jm\phi}, e^{jm\phi}, \sin(m\phi), \cos(m\phi)
	\label{eq:phi_terms}
\end{equation}
\begin{equation}
	R(\rho) \sim J_m(k_\rho\rho), Y_m(k_\rho\rho), H_m^{(1)}(k_\rho\rho), H_m^{(2)}(k_\rho\rho)
	\label{eq:r_terms}.
\end{equation}

\begin{figure}
	\includegraphics[width=0.48\textwidth]{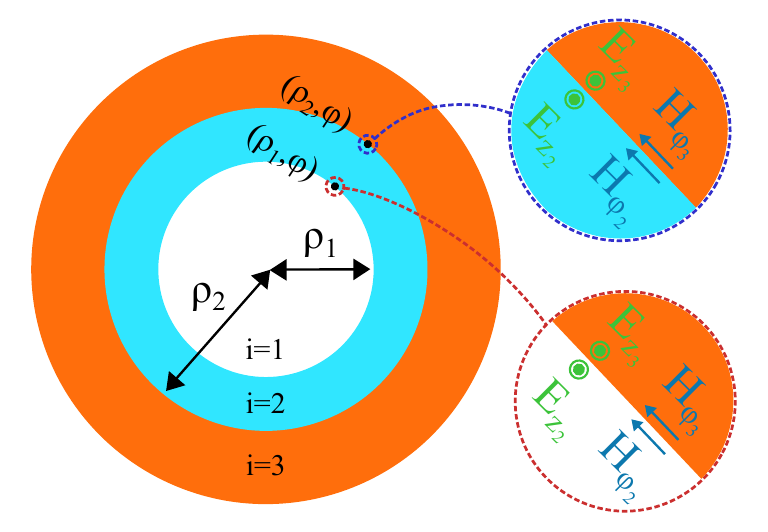}
	\caption{Tangential electric and magnetic fields of the axial symmetric ${TM}_{on}$-modes at the two interfaces of the DLW with vacuum at radius $\rho_1$ and the dielectric with the metal at $\rho_2$.}
	\label{fig:Boundary}
\end{figure}

Now we are able to compute the electric and magnetic fields of TM modes by using Eq.~(\ref{eq:Electric and magnetic field from potential}) from the magnetic vector potential $\vec{A}$ as follows \cite{fields1961time}.
\begin{equation}
	\label{eq:Fields from Potential}
	\begin{alignedat}{2}
		E_\rho &= \frac{1}{j\omega\varepsilon}\frac{\partial^2\Psi}{\partial\rho\partial z} &&\qquad H_\rho = \frac{1}{\rho}\frac{\partial\Psi}{\partial\phi} \\
		E_\phi &= \frac{1}{j\omega\varepsilon\rho}\frac{\partial^2\Psi}{\partial\phi\partial z} &&\qquad H_\phi = -\frac{\partial\Psi}{\partial\rho} \\
		E_z &= \frac{1}{j\omega\varepsilon}\left(\frac{\partial^2}{{\partial z}^2}+k^2\right)\Psi &&\qquad H_z = 0
	\end{alignedat}
\end{equation}
For the $TM$ modes of concentric cylindrical multilayer structures which exhibit axial symmetry (\(\frac{\partial\psi}{\partial\phi}=0, m=0\)), the electric and magnetic fields of the \(i\)-th layer (see Fig.~\ref{fig:Boundary}) are formulated as:
\begin{widetext}
\begin{equation}
	\begin{aligned}
		\label{eq:fields}	
		&E_z(\rho,z) = \left[A_i\begin{Bmatrix}
			J_0(k_{\rho_i}\rho)\\
			H_0^{(1)}(k_{\rho_i}\rho)\\
		\end{Bmatrix}
		+B_i\begin{Bmatrix}
			Y_0(k_{\rho_i}\rho)\\
			H_0^{(2)}(k_{\rho_i}\rho)
		\end{Bmatrix}\right]
			e^{-jk_zz} \qquad &&E_\phi(\rho,z) = 0  \\
		&E_\rho(\rho,z) = -\frac{jk_z}{k_{\rho_i}}\left[A_i
		\begin{Bmatrix}
			J_1(k_{\rho_i}\rho)\\
			H_1^{(1)}(k_{\rho_i}\rho)
		\end{Bmatrix}
		+B_i
		\begin{Bmatrix}
			Y_1(k_{\rho_i}\rho)\\
			H_1^{(2)}(k_{\rho_i}\rho)
		\end{Bmatrix}\right]	
			e^{-jk_zz} \qquad &&H_\rho(\rho,z) = 0  \\
		&H_\phi(\rho,z) = -\frac{j\omega\varepsilon_{d_i}}{k_{\rho_i}}\left[A_i
		\begin{Bmatrix}
			J_1(k_{\rho_i}\rho)\\
			H_1^{(1)}(k_{\rho_i}\rho)
		\end{Bmatrix}
		+B_i
		\begin{Bmatrix}
			Y_1(k_{\rho_i}\rho)\\
			H_1^{(2)}(k_{\rho_i}\rho)
		\end{Bmatrix}\right]	
		e^{-jk_zz} \qquad &&H_z(\rho,z) = 0  \\
	\end{aligned}
\end{equation}
\end{widetext}
where the Hankel functions are used for the outer layer and Bessel and Neumann functions are used for the two inner layers. The coefficient of the Neumann function $Y_0\left(k_{\rho_1}\rho\right)$ for the first layer is zero since the Neumann functions have a singularity at the origin. To account for the behavior of the outer layer, only the first kind (or second kind depends on the definition) of Hankel function $H_0^{\left(1\right)}\left(k_{\rho_3}\rho\right)$ is utilized since the fields need to approach zero as the radius $\rho$ goes to infinity.  

In general, $k_z$ can be a complex number when dealing with lossy structures. To establish the relationship among the unknown variables, it is essential to write the continuity equations at the interfaces. Tangential electric and magnetic fields must be continuous functions of $\rho$ at both interfaces for any angle $\phi$ (Fig.~\ref{fig:Boundary}).

For a three-layer structure with a vacuum core, expressing the matrix form of the continuity of electric and magnetic fields at both interfaces results in Eq.~(\ref{eq:Matrix4x4}).
\begin{widetext}
\begin{equation}
	\label{eq:Matrix4x4}	
	\begin{gathered}
	\left[\begin{matrix}
		J_0(k_{\rho1}\rho_1) & -J_0(k_{\rho2}\rho_1) & -Y_0(k_{\rho2}\rho_1) & 0 \\[0.2cm]
		\frac{{\varepsilon_1}J_1(k_{\rho1}\rho_1)}{k_{\rho1}} & -\frac{{\varepsilon_2}J_1(k_{\rho2}\rho_1)}{k_{\rho2}} & -\frac{{\varepsilon_2}Y_1(k_{\rho2}\rho_1)}{k_{\rho2}} & 0 \\[0.2cm]
		0 & -J_0(k_{\rho2}\rho_2) & -Y_0(k_{\rho2}\rho_2) & H_0^{(1)}(k_{\rho3}\rho_2) \\[0.2cm]
		0 & -\frac{{\varepsilon_2}J_1(k_{\rho2}\rho_2)}{k_{\rho2}} & -\frac{{\varepsilon_2}Y_1(k_{\rho2}\rho_2)}{k_{\rho2}} & \frac{{\varepsilon_3}H_1^{(1)}(k_{\rho3}\rho_2)}{k_{\rho3}}
	\end{matrix}\right] 
	\left[\begin{matrix}
	A_1 \\
	A_2 \\
	B_2 \\
	A_3
	\end{matrix}\right] 
	=
	\left[\begin{matrix}
	0 \\
	0 \\
	0 \\
	0
	\end{matrix}\right]
	\end{gathered}
\end{equation}
\end{widetext}	

In order to satisfy Eq.~(\ref{eq:Matrix4x4}), the determinant of the coefficient matrix is set to zero (characteristic equation), resulting in nontrivial solutions for the fields. 
Now the only unknown variable remaining in the characteristic equation is the longitudinal component of the wave vector $k_z$ that can be determined using numerical techniques. Consequently, we can find the electromagnetic field distribution of the ${TM}_{01}$-mode in the DLW and important properties such as the phase velocity, group velocity, and attenuation coefficient, providing valuable insights into the behavior of the wave propagation within the structure.
It's important to acknowledge that the aim of this paper is to provide a general exploration of DLW design principles. However, to quantitatively demonstrate the influence of various parameters and graphically show the design procedure specific values for some parameters must be chosen. Table~\ref{tab:DLW parameters} lists the parameters assumed in solving the characteristic equation of the DLW. Please note that we collectively define and optimize all the parameters in the third section.
\begin{table}
	\caption{\label{tab:DLW parameters}DLW parameters}
	\begin{ruledtabular}
		\begin{tabular}{ll}
			{\textrm{\textbf{Parameter}}} & {\textrm{\textbf{Value}}}\\
			\colrule
			{\textrm{\textbf{Frequency}}} & 300~GHz\\

			{\textrm{\textbf{DLW dimensions}}}&\\
			\qquad Vacuum radius & \SI{200}{\micro\meter}\\
			\qquad Dielectric thickness & \SI{143.1}{\micro\meter}\\

			\textrm{\textbf{Material properties}} & \\ 
			\qquad \textrm{Conductivity of metal} & 5.9e7~S/m\\
			\qquad Dielectric refractive index & 1.95\\
			\qquad Dielectric attenuation constant (300~GHz) & $\sim$10~Np/m\\
		\end{tabular}
	\end{ruledtabular}
\end{table}

\begin{figure}
	\includegraphics[width=0.8\linewidth]{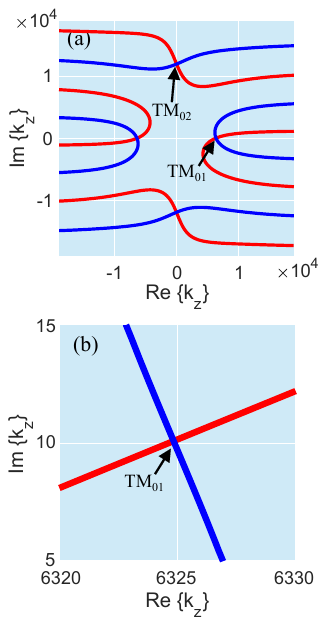}
	\caption{Loci of the zeros of real and imaginary part of the coefficient matrix determinant in the complex $k_z$-plane. (b) High resolution graph near one of the complex zeros in the $k_z$-plane. 
	}
	\label{fig:zeros of Matrix}
\end{figure}

Fig.~\ref{fig:zeros of Matrix} displays contours representing the loci of the zeros of the real and imaginary parts of the determinant (characteristic equation) as a function of the complex values of $k_z$ for the DLW (in order to avoid any singularity caused by $k_{\rho_2}$ which is the denominator of some elements of the coefficient matrix, we multiply the second and forth rows of the matrix by $k_{\rho_2}$ and calculate the determinant afterwards).
\begin{figure*}
	\centering
	\includegraphics[width=0.9\textwidth]{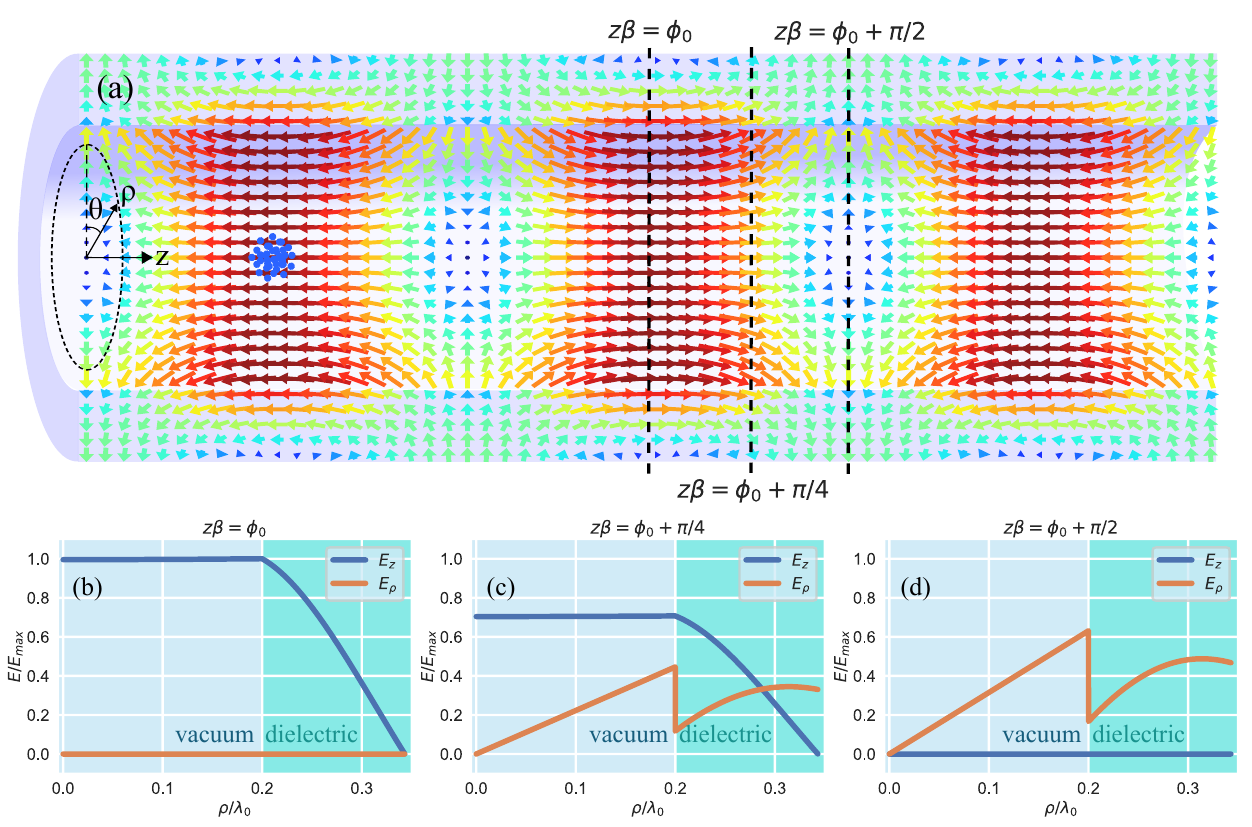}
	\caption{(a) Electric field distribution of the ${TM}_{01}$-mode within the DLW with parameters listed in Table~\ref{tab:DLW parameters} for a cross section along the waveguide at a fixed time t. The filed distribution is radially symmetric around the propagation axis. (b-d) shows the fields over the waveguide cross-section at the different propagation phases indicated in Figure (a). The dielectric thickness-to-vacuum radius ratio has been deliberately selected to align the phase velocity with the speed of light (listed in Table~\ref{tab:DLW parameters}).}
	\label{fig:Fild distribution}
\end{figure*}

As seen in Fig.~\ref{fig:zeros of Matrix}, Eq.~(\ref{eq:Matrix4x4}) does not have only one solution for $k_z$, each solution corresponds to one of the ${TM}_{0n}$ modes. The ${TM}_{01}$ mode is the first propagating TM mode among the solutions inside the DLW which has the highest real value for $k_z$.  

Subsequently, the electric field distribution is calculated using Eq.~(\ref{eq:fields}) for the ${TM}_{01}$-mode within the DLW and it is illustrated in Fig.~\ref{fig:Fild distribution}.

As depicted in Fig.~\ref{fig:Fild distribution}, the z-component of the electric field $E_z$ exhibits complete uniformity across the entire vacuum aperture ($k_{\rho_1}={k_0}^2-{k_z}^2\cong0$) whenever the phase velocity is set to the speed of light ($\frac{\omega}{\Re\{k_z\}}=\frac{\omega}{k_0}$). Notably, in case of a lossy structure ($\Im\{{k}_z\}\neq0$),\ $k_{\rho_1}$ does not vanish anymore resulting in a small nonuniformity of $E_z$. However, since the attenuation level is not high here ($\Re\{{k}_z\}\gg \Im\{{k}_z\}$), the $k_{\rho_1}$ value is sufficiently low and does not change the uniformity of $E_z$ in the vacuum area considerably. With these assumptions, the terms $\frac{J_1\left(k_{\rho_1}\rho\right)}{k_{\rho_1}}$ and $J_0\left(k_{\rho_1}\rho\right)$ in Eq.~(\ref{eq:fields}) approach $\frac{\rho}{2}$ and $1$, respectively, resulting in a simple relation between the field components. As a consequence, the radial and axial (longitudinal) electric fields in the vacuum region are related by
\begin{equation}
	\label{eq:Ep/Ez}
	\frac{E_\rho}{E_z}\cong-\frac{j\pi\rho}{\lambda}
\end{equation}
 as the phase velocity equals to the speed of light. As inferred from Eq.~(\ref{eq:Ep/Ez}) the radial and longitudinal components of the electric field are $\frac{\pi}{2}$ out of phase. At the peak acceleration phase, as shown in Fig.~\ref{fig:Fild distribution}(b), the radial electric field is nearly zero and gradually increases over time and is always a linear function of $\rho$. As a consequence of Eq.~(\ref{eq:Ep/Ez}), the radial electric field component rises to approximately 63\% of the maximum longitudinal electric field magnitude after one-quarter of a period at the dielectric interface for a DLW with 200 µm vacuum radius at 300 GHz operating frequency  (Fig.~\ref{fig:Fild distribution}(d)).  
The angular component of the magnetic field can also be expressed with the help of radial component of electric field by Eq.~(\ref{eq:fields}) ($H_\phi=\frac{{\omega\varepsilon}_{d_i}}{k_z}E_\rho$). Given that the imaginary part of the term $\frac{\varepsilon_{d_i}}{k_z}$ is   negligible compared to its real part, $H_\phi$ is an in-phase linear function  of $E_\rho$. This implies that when the electron experiences the peak acceleration field (Fig.~\ref{fig:Fild distribution}(b)) the magnetic field does not affect the electron’s motion. But as the electron approaches phases with zero longitudinal electric field, the force due to the magnetic field increases as a linear function of  radius. 
It is worth noting that the magnetic field has no effect on the electrons positioned on the axis of the DLW at any given phase.   

\subsection{\label{sec:1.2	Phase velocity and attenuation coefficient}Phase velocity and attenuation coefficient}

\begin{figure}
	\includegraphics[width=.9\linewidth]{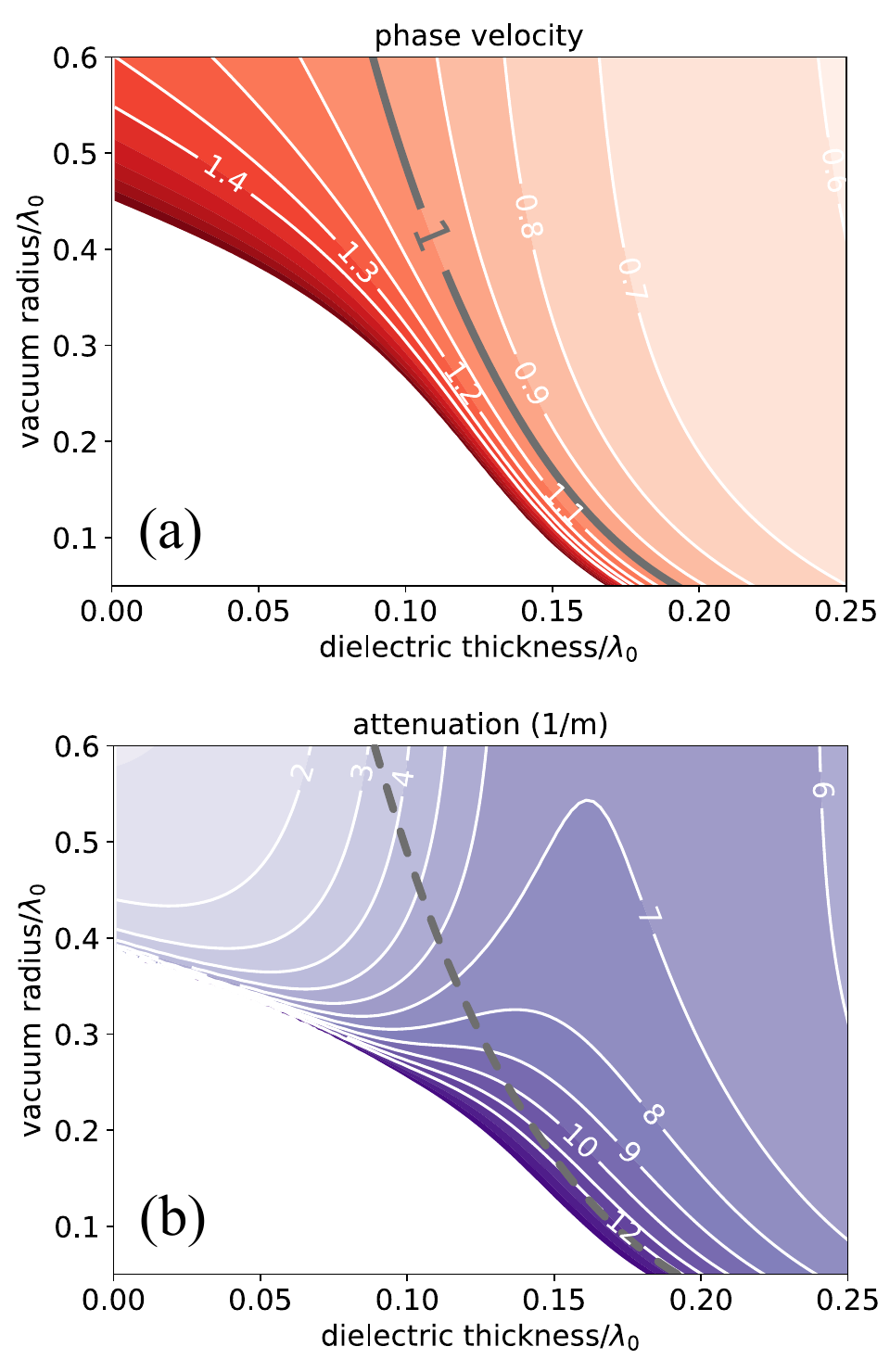}
	\caption{(a) Phase velocity normalized to the vacuum speed of light, and (b) negative imaginary part of propagation constant, $\alpha=-\ \Im \left\{k_z\right\}$, i.e. attenuation coefficient, of the ${TM}_{01}$-mode of the DLW versus vacuum radius and dielectric thickness normalized to vacuum wavelength (since the phase velocity is independent of wavelength for a given complex refractive indices of the dielectric $\varepsilon_2$ and metal $\varepsilon_3$). Here we assumed $\varepsilon_{r_2}=3.8+j0.0016$ and $\sigma_3=5.96e7~S/m$ at 300GHz.}
	\label{fig:vph and alpha}
\end{figure}

After determining the complex propagation constant $k_z$, the phase velocity, group velocity, and attenuation coefficient can be derived from Eqs.~(\ref{eq:vph}-\ref{eq:PC AC}).
\begin{equation}
	\label{eq:vph}
	v_{ph}=\frac{\omega}{\beta}
\end{equation}
\begin{equation}
	\label{eq:vg}
	v_g=\frac{\partial\omega}{\partial\beta}
\end{equation}
\begin{equation}
	\label{eq:PC AC}
	\beta=\Re\{{k}_z\}, \alpha=-\Im\{{k}_z\}
\end{equation}
Fig.~\ref{fig:vph and alpha} illustrates normalized phase velocity (normalization with respect to vacuum speed of light), and attenuation coefficient of the ${TM}_{01}$-mode of the DLW with a dielectric material having a refractive index of $n_2=1.95$ (fused Silica) and absorption coefficient of $10$~(Np/m) at 300~GHz \cite{bagdad1968far,tsuzuki2015influence,koike1989optical,naftaly2021terahertz,afsar1984millimeter,dutta1986complex}, corresponding to $\varepsilon_{r_2}=3.8+j0.0016$, which is surrounded by a copper layer with a conductivity of $\sigma=5.9e7$~(S/m), corresponding to $\varepsilon_{r_3}=1+\ j3.6e6$ at the same frequency. Since the refractive index of the dielectric (fused silica) is roughly uniform for a wide range of frequencies below 500~GHz, the calculated phase velocity values are expandable to other frequencies. But since the attenuation coefficient varies by frequency, the results for the attenuation are valid only for 300~GHz. 

The dimensions of the DLW should be carefully chosen to ensure that the phase velocity is in a range greater than the initial velocity of electrons and less than the vacuum speed of light.
To maintain a specific phase velocity, reducing the vacuum radius will necessitate increasing the dielectric thickness (contours of Fig.~\ref{fig:vph and alpha}(a)), resulting in changes to the electric field along the axis. In the subsequent section, we will conduct calculations to determine the electric field  for different dimensions.

\subsection{\label{sec:Power, field, and group velocity}Power, field, and group velocity}
To determine the electric field over a cross section for a given drive power, we need to compute the total power transported by the ${TM}_{01}$-mode of the DLW in terms of the field amplitude. It should be noted that in this paper, we do not yet investigate the coupling to the DLW. Therefore, the terms “power” and “energy” refer to the power and energy transported by the ${TM}_{01}$-mode.

\begin{figure}
	\centering
	\includegraphics[width=.9\linewidth]{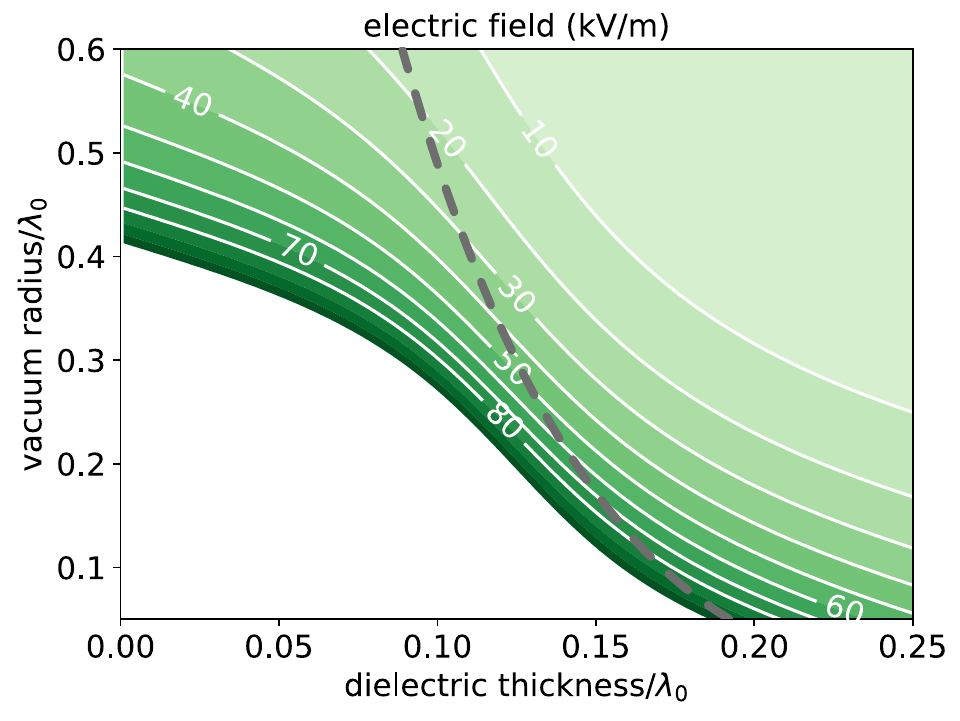}
	\caption{Longitudinal component of the electric field on the axis of the DLW considering a unit power of 1W for the parameters listed in Table~\ref{tab:DLW parameters}. The gray dashed line indicates the condition for phase velocity equal to the speed of light.}
	\label{fig:Ez}
\end{figure}

In order to calculate the average power (for a sinusoidally varying field) through an area $s$ with unit normal vector $\hat{n}$, one needs to compute the time-averaged integral of the inner product of the Poynting vector $\vec{S}$ and the normal vector $\hat{n}$ over the cross-sectional area.
\begin{equation}
	\label{eq:Ps}
	P_s=\frac{1}{2}\Re\left\{\iint_{s}{\vec{S}\cdot\hat{n}\ ds}\right\}
\end{equation}
where $\vec{S}$ is defined by Eq.~(\ref{eq:S}) for complex electric and magnetic fields $\vec{E}$ and $\vec{H}$.
\begin{equation}
	\label{eq:S}
	\vec{S}=\frac{1}{2}\vec{E}\times{\vec{H}}^\ast
\end{equation}
Consequently, the total longitudinal power $P_z$ flowing through a cross section of the DLW is computed by integrating the z-component of the Poynting vector $S_z$ over the entire cross-sectional area. For the given three-layer structure, the relationship can be expressed by Eq.~(\ref{eq:PzTotal}).
\begin{widetext}
\begin{equation}
	\begin{split}
		\label{eq:PzTotal}	
		P_z=P_{z_1}+P_{z_2}+P_{z_3}=
		\int_{0}^{\rho_1}{\int_{0}^{2\pi}{S_z\left(A_1\right)\ \rho d\theta d\rho}+\int_{\rho_1}^{\rho_2}\int_{0}^{2\pi}{S_z\left(A_2,B_2\right)\ \rho d\theta d\rho}+\int_{\rho_2}^{\infty}\int_{0}^{2\pi}{S_z\left(A_3\right)\ \rho d\theta d\rho}}
	\end{split}
\end{equation}
\end{widetext}
The coefficients $A_2$, $B_2$, and $A_3$ could be written as functions of $A_1$ utilizing Eq.~(\ref{eq:Matrix4x4}). By equating the right side of Eq.~(\ref{eq:PzTotal}) to the given power, it becomes possible to calculate the coefficients $A_1$, $A_2$, $B_2$, $A_3$ and consequently, determine the absolute value of the electromagnetic field within the DLW where the axial electric field $E_z$ is the primary field component responsible for acceleration. Fig.~\ref{fig:Ez} illustrates the axial electric field of the cylindrical DLW for a unit power of 1~W.

\begin{figure}[ht]
	\centering
	\includegraphics[width=.9\linewidth]{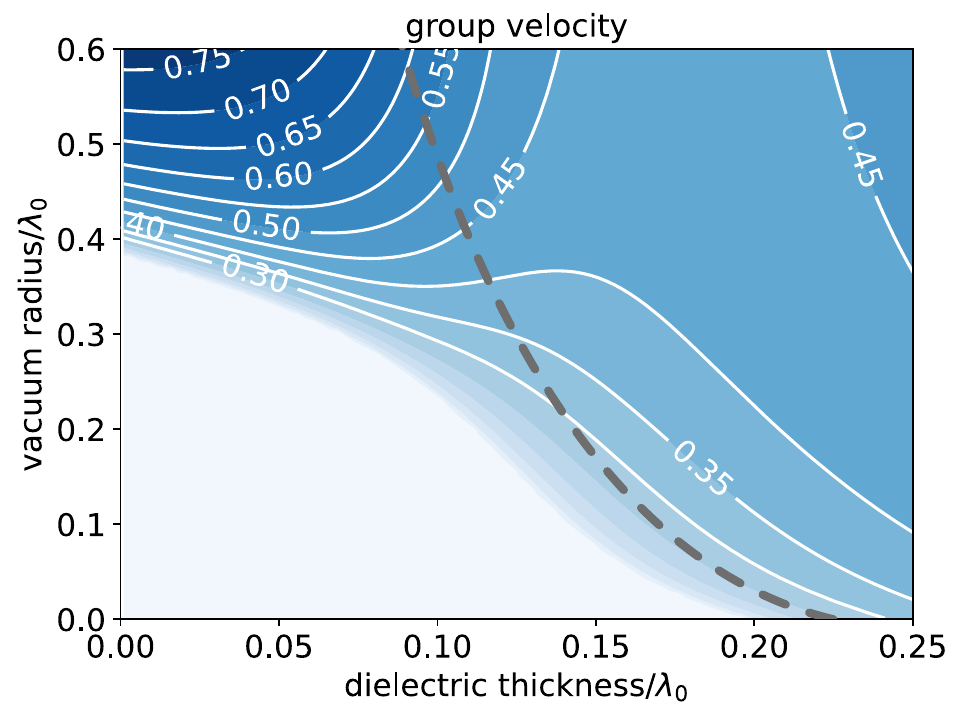}
	\caption{Group velocity of the ${TM}_{01}$-mode normalized to the vacuum speed of light for the parameters listed in Table~\ref{tab:DLW parameters}. The gray dashed line indicates the condition for phase velocity equal to the speed of light.}
	\label{fig:vg}
\end{figure}

Considering that the power flow in the layers has been computed to determine the absolute field values, we can employ a simpler method than Eq.~(\ref{eq:vg}) to calculate the group velocity. The connection between group and phase velocity, as revealed through axial power flow within the structure, is examined in \cite{kawakami1975relation}. The relationship between $v_{ph}$ and $v_g$ relative to $c$ (vacuum speed of light) can be described as
\begin{equation}
	\label{eq:}	
	\frac{\nu_{ph}\nu_g}{c^2}=\frac{\int_{s} S_zds}{\int_{s}{n^2S}_zds}
\end{equation}
Here, $s$ represents the entire cross-section of the waveguide. We have calculated phase velocity $v_{ph}$, power flow in the vacuum layer $P_{z_1}$, and power flow in the dielectric layer $P_{z_2}$ by Eq.~(\ref{eq:PzTotal}). Consequently, the normalized group velocity can be expressed as
\begin{equation}
	\label{eq:vg from power flow}	
	\frac{v_g}{c}=\frac{c\ \left(P_{z_1}+P_{z_2}\right)}{v_{ph}\left(P_{z_1}+\varepsilon_2P_{z_2}\right)}	
\end{equation}
We note  that the power flow through metal $P_{z3}$ is neglected. Fig.~\ref{fig:vg} shows the normalized group velocity of the ${TM}_{01}$-mode. As can be derived from Eq.~(\ref{eq:vg from power flow}), an increase in the power flow through the dielectric region leads to a reduction in group velocity. The minimum group velocity of the ${TM}_{01}$-mode (for normalized phase velocity of 1) is $\frac{1}{\varepsilon_2}$ , which is approached for very small vacuum radii.
As depicted in Fig.~\ref{fig:vg}, the group velocity is much lower than the speed of light. Evidently, achieving larger group velocity values is desirable as it allows the electron to interact with the THz pulse over a longer distance. However, obtaining higher group velocities necessitates increasing the vacuum radius, which, in turn, leads to a reduced accelerating field for a given power. A previous study introduced an upper limit for the group velocity based on the axial and transverse electric field strengths at the edges of the vacuum region \cite{england2014dielectric}.
\begin{equation}
	\label{eq:vg_Ez}    
	\nu_g \leq \frac{c}{1+2\left(\frac{E_z}{|\pmb{E}_\bot|}\right)^2}
\end{equation}
where $\pmb{E}_\bot$ represents the transverse electric field, equivalent to $E_\rho$ for the ${TM}_{01}$-mode. As evident from Eq.~(\ref{eq:vg_Ez}), increasing the group velocity will result in a reduced accelerating field $E_z$.

\begin{figure}
	\centering
	\includegraphics[width=.9\linewidth]{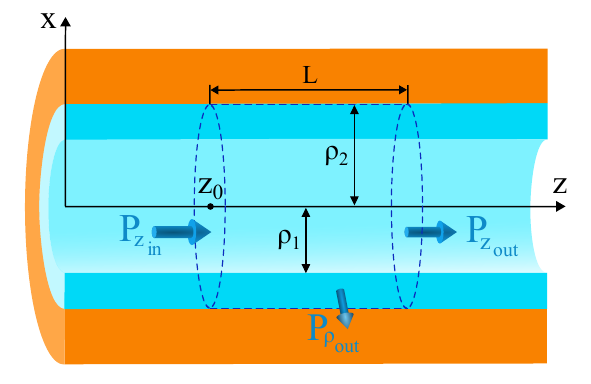}
	\caption{Power flows for a cylinder with the radius of $\rho_2$. $P_{\rho_{out}}$ is the outward radial power flow over a length L. $P_{z_{in}}$ and $P_{z_{out}}$ are longitudinal powers flowing into and out of the cylinder.}
	\label{fig:Poynting vector}
\end{figure}

Up to this point, we have employed the Poynting vector to determine the accelerating field and group velocity. Now, we want to utilize it to calculate the power attenuation within both the dielectric and metallic layers. The attenuation values calculated in the preceding section accounted for losses in the entire structure, including both dielectric and metal losses. To determine the individual contribution of each layer to the overall attenuation, one could integrate the dissipated power over the volume. We choose to calculate the dissipated power within the dielectric layer using the Poynting theorem. We consider  a cylinder with radius $\rho_2$ as depicted in Fig.~\ref{fig:Poynting vector}, where the term $P_{\rho_{out}}$ denotes the radial power flow through the dielectric-metal interface over a length $L$. $P_{z_{in}}$ and $P_{z_{out}}$ refer to the axial power flow entering and exiting the cylinder, respectively.

The conservation relation for the power flows and dissipated power within the cylinder in Fig.~\ref{fig:Poynting vector} can be expressed as
\begin{equation}
	\label{eq:} 
	P_{z_{in}}\left(\rho\right)=P_{\rho_{out}}\left(\rho\right)+P_{z_{out}}\left(\rho\right)+P_{d_{dielctric}}
\end{equation}
where $P_{d\ dielctric}$ is the dissipated power within the dielectric layer. By utilizing Eqs.~(\ref{eq:fields}), (\ref{eq:Ps}), and (\ref{eq:S}) the output power flow $P_{z_{out}}$ can be expressed as $e^{-2\alpha L}P_{z_{in}}$ since all the fields are attenuated with the attenuation constant $\alpha$. Thus, the dissipated power within the dielectric is calculated as
\begin{widetext}
\begin{equation}
	\begin{split}
		\label{eq:} 
		P_{d_{\text{dielectric}}} &= \left(1-e^{-2\alpha L}\right)P_{z_{\text{in}}} - \int_{z_0}^{z_0+L}\int_{0}^{2\pi} \frac{1}{2}\operatorname{Re}\left\{E_z(\rho_2,z_0)H_\phi^*(\rho_2,z_0)\ e^{-2\alpha(z-z_0)}\rho_2d\theta dz\right\} \\
		&= \left(1-e^{-2\alpha L}\right)\left(P_{z_{\text{in}}}-\frac{\pi\rho \operatorname{Re} \left\{E_z(\rho_2,z_0)H_\phi^*(\rho_2,z_0)\right\}}{2\alpha}\right)
	\end{split}
\end{equation}
\end{widetext}

Notably, due to axial symmetry of the electric and magnetic fields for the ${TM}_{01}$-mode, the angular integration term, is replaced by $2\pi$. 
The total power loss $P_{d_{total}}$ could be formulated using the same method assuming a cylinder with a sufficiently large diameter. If the skin depth of the metallic layer is very short in comparison with the dimension of the DLW, the longitudinal power flow within the metallic layer is accordingly negligible in comparison with the power flow inside the dielectric and vacuum ($P_{z_3}\ll P_{z_1}+P_{z_2}$). Consequently, the total dissipated power can be written as
\begin{equation}
	\label{eq:Pd_total} 
	P_{d_{total}}\cong P_{z_{out}}-P_{z_{in}}=\left(1-e^{-2\alpha L}\right)P_{z_{in}}
\end{equation} 
$P_{z_{in}}$ is determined by integrating the z-component of the Poynting vector over the flat surface of the cylinder (Fig.~\ref{fig:Poynting vector}) as utilized for calculating the power flow in Eq.~(\ref{eq:PzTotal}). Therefore $P_{d_{total}}$ is
\begin{equation}
	\label{eq:} 
	P_{d_{\text{total}}} \cong \left(1-e^{-2\alpha L}\right)\pi\int_{0}^{\rho_2} \operatorname{Re}\left\{E_\rho(\rho)H_\phi^*(\rho)\right\}\rho\, d\rho
\end{equation}

The ratio of the dissipated power inside the metallic and dielectric layers to the total dissipated power could be formulated as
\begin{subequations}
\begin{equation}
	\label{eq:} 
		\frac{P_{d_{\text{dielectric}}}}{P_{d_{\text{total}}}} \cong 1 - \frac{\rho_2 \operatorname{Re} \left\{E_z(\rho_2,z_0)H_\phi^*(\rho_2,z_0)\right\}}{2\alpha\int_{0}^{\rho_2} \operatorname{Re}\left\{E_\rho(\rho)H_\phi^*(\rho)\right\}\rho\, d\rho} 
\end{equation}
\begin{equation}
		\frac{P_{d_{\text{metal}}}}{P_{d_{\text{total}}}} \cong \frac{\rho_2 \operatorname{Re} \left\{E_z(\rho_2,z_0)H_\phi^*(\rho_2,z_0)\right\}}{2\alpha\int_{0}^{\rho_2} \operatorname{Re}\left\{E_\rho(\rho)H_\phi^*(\rho)\right\}\rho\, d\rho}
\end{equation}
\end{subequations}
In order to see the effect of the metal loss on the dissipated power, different materials are examined. Table~\ref{tab:conductivity} shows the conductivity of some metals and its approximate equivalent skin depth at 300~GHz (skin depths are calculated with low frequency conductivity values \cite{serway1998principles,moore1967physical,bel2007electrical}). Although, the conductivity values are valid for a wide range of frequencies from DC to THz, one can employ the classical Drude model for more precise calculations \cite{liu20161,smith2001classical}.

\begin{table}
	\caption{\label{tab:conductivity}Conductivity and skin depth for various metals at 300~GHz}
	\begin{ruledtabular}
	\begin{tabular}{lcc}
	\textrm{Material} & \textrm{Conductivity (S/m)} & \textrm{Skin depth (nm)} \\
	\colrule
	Silver & $6.2 \times 10^7$ & $\sim 116$ \\
	Copper & $5.9 \times 10^7$ & $\sim 119$ \\
	Gold & $4.5 \times 10^7$ & $\sim 137$ \\
	Chromium & $7.9 \times 10^6$ & $\sim 326$ \\
	Titanium & $2.5 \times 10^6$ & $\sim 580$ \\
\end{tabular}
	\end{ruledtabular}
\end{table}

Table~\ref{tab:Dielectric attenuation and metal attenuation comparison for various vacuum radii} presents the total attenuation coefficient $\alpha$ for the ${TM}_{01}$-mode of the DLW and relative dissipated power within each layer for various metals and three different vacuum radii at the frequency of 300~GHz. The dielectric layer is composed of fused silica with the refractive index of 1.95 and attenuation coefficient 10~Np/m at 300~GHz \cite{afsar1984millimeter}. The dielectric thickness is optimized for each vacuum radius to set the phase velocity to the speed of light.

\begin{table*}
	\centering
	\caption{Dielectric attenuation and metal attenuation comparison for various vacuum radii}
	\label{tab:Dielectric attenuation and metal attenuation comparison for various vacuum radii}
	\begin{tabular}{cccccccccc}
		\toprule
		\rowcolor{gray!30} \cellcolor{gray!0} & \multicolumn{3}{c}{\textbf{Total Attenuation}} & \multicolumn{3}{c}{\cellcolor{gray!50}\textbf{Dielectric Attenuation}} & \multicolumn{3}{c}{\textbf{Metal Attenuation}} \\
		\rowcolor{gray!30} \cellcolor{gray!0} & \multicolumn{3}{c}{(\si{\per\meter})} & \multicolumn{3}{c}{\cellcolor{gray!50}to Total Attenuation} & \multicolumn{3}{c}{to Total Attenuation} \\
		\cmidrule(lr){2-4} \cmidrule(lr){5-7} \cmidrule(lr){8-10}
		\rowcolor{gray!30} \multicolumn{1}{c}{\textbf{\makecell{Vacuum radius}}} & \textbf{150} \si{\micro\meter} & \textbf{200} \si{\micro\meter} & \textbf{250} \si{\micro\meter} & \multicolumn{1}{c}{\cellcolor{gray!50}\textbf{150} \si{\micro\meter}} & \multicolumn{1}{c}{\cellcolor{gray!50}\textbf{200} \si{\micro\meter}} & \multicolumn{1}{c}{\cellcolor{gray!50}\textbf{250} \si{\micro\meter}} & \multicolumn{1}{c}{\cellcolor{gray!30}\textbf{150} \si{\micro\meter}} & \multicolumn{1}{c}{\cellcolor{gray!30}\textbf{200} \si{\micro\meter}} & \multicolumn{1}{c}{\cellcolor{gray!30}\textbf{250} \si{\micro\meter}} \\
		& & & & & & & & & \\
		\rowcolor{gray!30} \multicolumn{1}{c}{\textbf{\diagbox[width=15em, height=5em]{\qquad \qquad \large Material}{\makecell{Dielectric thickness}}}} & 154.4 \si{\micro\meter} & 143.1 \si{\micro\meter} & 132.8 \si{\micro\meter} & \cellcolor{gray!50} 154.4 \si{\micro\meter} &\cellcolor{gray!50} 143.1 \si{\micro\meter} &\cellcolor{gray!50} 132.8 \si{\micro\meter} & 154.4 \si{\micro\meter} & 143.1 \si{\micro\meter} & 132.8 \si{\micro\meter} \\
		& & & & & & & & & \\	
		Silver & 10.8 & 9.9 & 8.8 & \cellcolor{gray!20} 37 \% &\cellcolor{gray!20} 35 \% &\cellcolor{gray!20} 32 \% & 63 \% & 65 \% & 68 \% \\
		Copper & 11.0 & 10.1 & 9.0 &\cellcolor{gray!20} 37 \% &\cellcolor{gray!20} 34 \% &\cellcolor{gray!20} 32 \% & 63 \% & 66 \% & 68 \% \\
		Gold & 12.0 & 11.0 & 9.9 &\cellcolor{gray!20} 34 \% &\cellcolor{gray!20} 31 \% & \cellcolor{gray!20}29 \% & 66 \% & 69 \% & 71 \% \\
		Chromium & 23.1 & 21.7 & 19.7 &\cellcolor{gray!20} 18 \% &\cellcolor{gray!20} 16 \% &\cellcolor{gray!20} 14 \% & 82 \% & 84 \% & 86 \% \\
		Titanium & 37.9 & 35.8 & 32.7 &\cellcolor{gray!20} 11 \% &\cellcolor{gray!20} 10 \% &\cellcolor{gray!20} 9 \% & 89 \% & 90 \% & 91 \% \\
		\bottomrule
	\end{tabular}
\end{table*}

As evident from the values in Table~\ref{tab:Dielectric attenuation and metal attenuation comparison for various vacuum radii}, the fractional attenuation in the dielectric remains relatively consistent irrespective of the metal. Altering the vacuum radius while keeping the phase velocity constant (at the speed of light) has a limited effect on the attenuation level. However, it significantly affects the axial electric field due to the substantial changes in cross-sectional area.

To explore the impact of changing the frequency, we present a comparison of the attenuation coefficient of the DLW at three frequencies: 200~GHz, 300~GHz, and 400~GHz in Table~\ref{tab:Dielectric attenuation and metal attenuation comparison for various frequencies} (at a fixed vacuum radius of 200~µm). The refractive index of the dielectric layer (fused silica) shows little variation within this frequency range, but the attenuation coefficient varies from $\sim$6~Np/m at 200~GHz to $\sim$18~Np/m at 400~GHz \cite{bagdad1968far,tsuzuki2015influence,koike1989optical,naftaly2021terahertz,afsar1984millimeter,dutta1986complex}.

\begin{table*}
	\centering
	\caption{Dielectric attenuation and metal attenuation comparison for various frequencies}
	\label{tab:Dielectric attenuation and metal attenuation comparison for various frequencies}
	\begin{tabular}{cccccccccc}
		\toprule
		\rowcolor{gray!30} \cellcolor{gray!0} & \multicolumn{3}{c}{\textbf{Total Attenuation}} & \multicolumn{3}{c}{\cellcolor{gray!50}\textbf{Dielectric Attenuation}} & \multicolumn{3}{c}{\textbf{Metal Attenuation}} \\
		\rowcolor{gray!30} \cellcolor{gray!0} & \multicolumn{3}{c}{(\si{\per\meter})} & \multicolumn{3}{c}{\cellcolor{gray!50}to Total Attenuation} & \multicolumn{3}{c}{to Total Attenuation} \\
		
		\cmidrule(lr){2-4} \cmidrule(lr){5-7} \cmidrule(lr){8-10}
		\rowcolor{gray!30} \multicolumn{1}{c}{\textbf{\makecell{\large Frequency}}} & \textbf{200} \si{GHz} & \textbf{300} \si{GHz} & \textbf{400} \si{GHz} & \multicolumn{1}{c}{\cellcolor{gray!50}\textbf{200} \si{GHz}} & \multicolumn{1}{c}{\cellcolor{gray!50}\textbf{300} \si{GHz}} & \multicolumn{1}{c}{\cellcolor{gray!50}\textbf{400} \si{GHz}} & \multicolumn{1}{c}{\cellcolor{gray!30}\textbf{200} \si{GHz}} & \multicolumn{1}{c}{\cellcolor{gray!30}\textbf{300} \si{GHz}} & \multicolumn{1}{c}{\cellcolor{gray!30}\textbf{400} \si{GHz}} \\
		& & & & & & & & & \\
		\rowcolor{gray!30} \multicolumn{1}{c}{\textbf{\diagbox[width=15em, height=5em]{\qquad \qquad \large Material}{\makecell{Dielectric thickness}}}} & 238.5 \si{\micro\meter} & 143.1 \si{\micro\meter} & 97.4 \si{\micro\meter} & \cellcolor{gray!50} 238.5 \si{\micro\meter} &\cellcolor{gray!50} 143.1 \si{\micro\meter} &\cellcolor{gray!50} 97.4 \si{\micro\meter} & 238.5 \si{\micro\meter} & 143.1 \si{\micro\meter} & 97.4 \si{\micro\meter} \\
		& & & & & & & & & \\

		Silver & 6.2 & 9.9 & 13.6 &\cellcolor{gray!20}  41 \% &\cellcolor{gray!20}  35 \% &\cellcolor{gray!20}  35 \% & 59 \% & 65 \% & 56 \% \\
		Copper & 6.3 & 10.1 & 13.9 &\cellcolor{gray!20}  40 \% &\cellcolor{gray!20}  34 \% &\cellcolor{gray!20}  34 \% & 60 \% & 66 \% & 66 \% \\
		Gold & 6.9 & 11.0 & 15.2 &\cellcolor{gray!20}  37 \% &\cellcolor{gray!20}  31 \% &\cellcolor{gray!20}  31 \% & 63 \% & 69 \% & 69 \% \\
		Chromium & 12.9 & 21.7 & 29.7 &\cellcolor{gray!20}  20 \% &\cellcolor{gray!20}  16 \% &\cellcolor{gray!20}  16 \% & 80 \% & 84 \% & 84 \% \\
		Titanium & 20.9 & 35.8 & 49.0 &\cellcolor{gray!20}  12 \% &\cellcolor{gray!20}  10 \% &\cellcolor{gray!20}  10 \% & 88 \% & 90 \% & 90 \% \\
		\bottomrule
	\end{tabular}
\end{table*}

Up to this point, we have explored the electromagnetic properties of the DLW. Our assumption is that the presence of a low-current electron beam does not significantly alter these electromagnetic fields. Neglecting the fields of the particles in the waveguide, i.e. wakefields, the computed field distributions enable us to solve the relativistic equation of motion of charged particles in the DLW.
In the following section, our focus will shift towards solving the equation of motion for an individual electron positioned on the axis of the DLW and determining its acceleration.

\section{\label{sec:Linear Acceleration}Linear acceleration}
The DLW LINAC is powered by a multicycle THz pulse traveling down the DLW. In the ideal case, the electron bunch travels together with the negative peak of the electric field. In case of having a nonrelativistic electron bunch, the velocity of the electrons should increase during the acceleration process while the THz pulse maintains a constant phase velocity (a uniform DLW in the acceleration direction assumed). Thus, the relative position of the electron and the THz wave changes as it undergoes acceleration. Consequently, the important aspect to consider is maintaining the electron bunch within a single half-cycle of the THz pulse to prevent deceleration. 

\begin{figure}
	\centering
	\includegraphics[width=.8\linewidth]{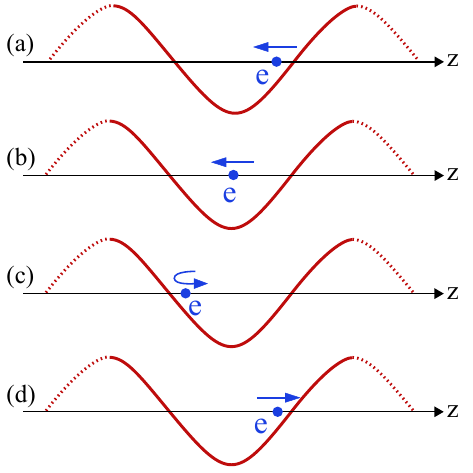}
	\caption{(a) Electron positioned at the leading part of the negative half-cycle during the initial stage of acceleration with a velocity lower than the phase velocity of the electron. (b) Electron experiencing maximum acceleration. (c) Electron reaching a velocity higher than the phase velocity of the THz pulse and drifting towards the leading part of the half-cycle. (d) Accelerated electron leaving the negative half-cycle and moving into a decelerating, positive half-cycle, which should be avoided. Blue arrows show the velocity of electron relative to the phase velocity of the negative half-cycle.}
	\label{fig:electron phase}
\end{figure}

The optimal scenario for an individual electron involves setting the phase velocity of the THz pulse higher than the initial velocity of the electron and lower than its final velocity ($\sim$1). The electron must be initially located in the leading portion of the negative electric field half-cycle in order to maximize the acceleration (optimal initial position is discussed later in this section). At the beginning, the electron starts to lag behind the front part of the negative half-cycle. However, as the electron accelerates, its velocity gradually approaches and eventually surpasses the phase velocity of the THz pulse. Therefore, it shifts from the lagging position to the leading position in the half-cycle, eventually leaving the accelerating half-cycle and entering the decelerating phase (see Fig.~\ref{fig:electron phase}). At this time the waveguide or THz field should end (if we are not constrained by a shorter pulse width). Another way to keep the electron in phase right after dephasing is to rephase the electric field using phase shifter structures \cite{zhang2022long}. Nevertheless, we will continue to focus on the axial uniform DLW.
To achieve an optimum design, the THz pulse width should be chosen based on the group velocity to avoid decelerating half-cycles. As the THz pulse propagates, the initially accelerating half-cycle transitions from the forefront rearward to the back end of the pulse due to the lower group velocity in comparison with the phase velocity. The length of the DLW and the pulse width are essential parameters in preventing electron deceleration.

We described the phase slippage of the electron within the DLW. Now we would like to quantitatively investigate the relative position of the electron with respect to the accelerating half-cycle, and the effect of the injection phase of the electron on the acceleration. 

\begin{figure}
	\centering
	\includegraphics[width=1\linewidth]{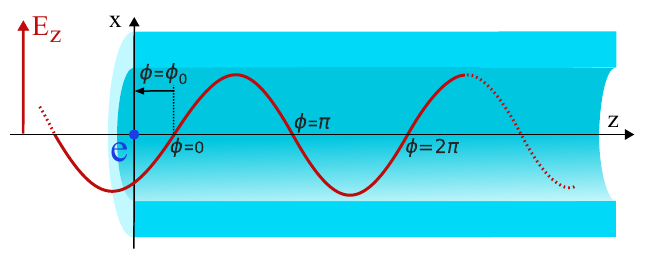}
	\caption{Relative phase of the electron with respect to the accelerating half-cycle.}
	\label{fig:injection phase_picture}
\end{figure}

We establish the relative phase of the electron with respect to the accelerating half-cycle as depicted in Fig.~\ref{fig:injection phase_picture}. Here, we designate $\phi_0$ as the relative phase at the moment the electron enters the DLW and call it electron injection phase. Accelerating half-cycles are characterized by phases within the range of $(2p-1)\pi<\phi<2p\pi$ while decelerating half-cycles fall within the range of $2p\pi<\phi<(2p+1)\pi$ ($p$ is an integer number).

\begin{figure}
	\centering
	\includegraphics[width=0.85\linewidth]{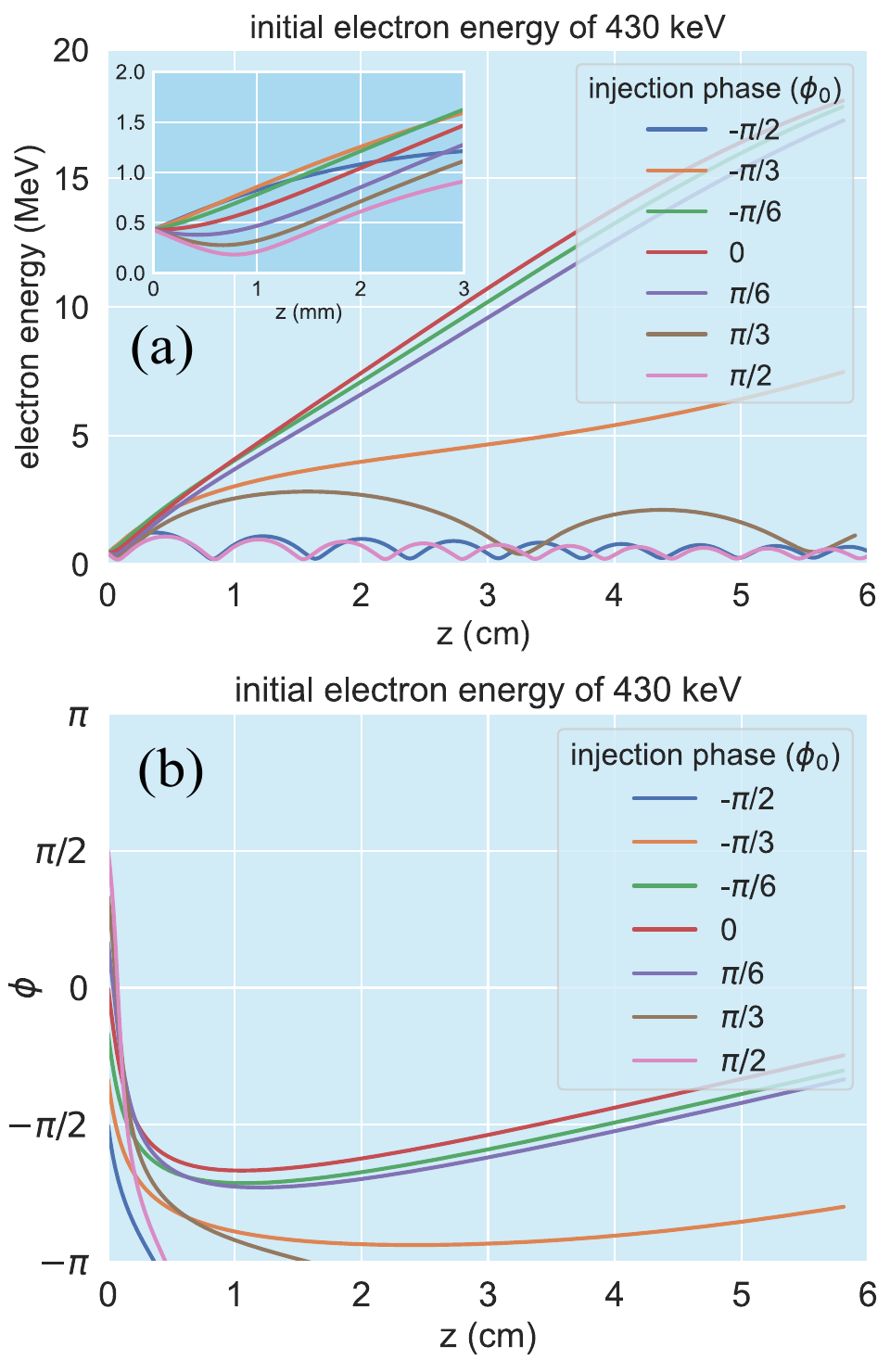}
	\caption{(a) Kinetic electron energy while being accelerated within the DLW and (b) its relative phase with respect to the accelerating half-cycle for different injection phases.}
	\label{fig:E_z_Phi_z_0_6}
\end{figure}

Optimal acceleration occurs when we maintain the relative phase of the electron $\phi$, between $0$ and $-\pi$ (injection phase assumed in the same range). Fig.~\ref{fig:E_z_Phi_z_0_6}(b) presents the variation in relative phase as the electron undergoes acceleration for different injection phase $\phi_0$ using the DLW and THz pulse parameters specified in Table~\ref{tab:DLW parameters} and Table~\ref{tab:THz pulse parameters}, respectively.
\begin{table}
	\caption{\label{tab:THz pulse parameters}THz pulse parameters}
	\begin{ruledtabular}
		\begin{tabular}{ll}
			{\textrm{\textbf{Parameter}}} & {\textrm{\textbf{Value}}}\\
			\colrule
			{\textrm{\textbf{THz pulse characteristics}}}&\\
			\qquad Frequency & \SI{300}{GHz}\\
			\qquad Energy & \SI{23}{mJ}\\
			\qquad Duration & \SI{467}{ps}
		\end{tabular}
	\end{ruledtabular}
\end{table}

Fig.~\ref{fig:E_z_Phi_z_0_6}(b) reveals that electrons injected with phases of $-\frac{\pi}{6},0,\ \frac{\pi}{6}$ have experienced notable acceleration compared to their counterparts. Remarkably, the electron injected with a positive phase of $\frac{\pi}{6}$, initially falling within a decelerating half-cycle rapidly transitions to an accelerating half-cycle within the first half of a millimeter of the DLW (as visible in the insets of  Fig.~\ref{fig:E_z_Phi_z_0_6}(a)). Consequently, the final energy of this electron is not significantly lower than that of others. 

\begin{figure}
	\centering
	\includegraphics[width=0.85\linewidth]{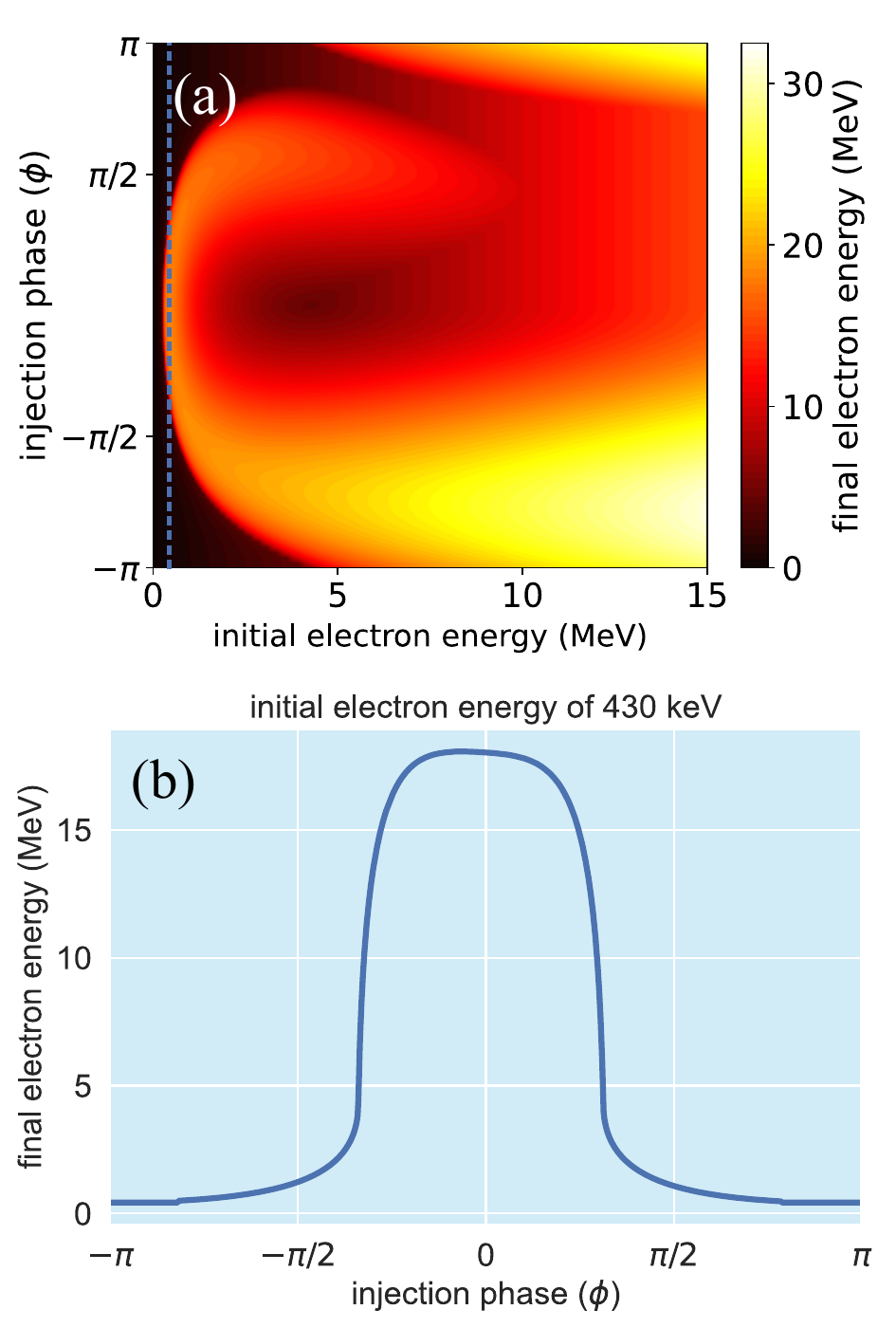}
	\caption{(a) Final kinetic electron energy for various injection phases and initial kinetic electron energy (b) Final kinetic electron energy on the dashed line in Fig.~(a) for a fixed initial electron energy of 430~keV.}
	\label{fig:energy injection phase sweep}
\end{figure}

Fig.~\ref{fig:energy injection phase sweep}(a), shows the final kinetic electron energy for different injection phases and initial kinetic electron energy. Increasing the initial kinetic electron energy results in a reduced phase slippage during the acceleration process.  This figure underscores the efficiency of injecting high energy electrons close to the peak of the electric field. Notably, for higher electron energy, a similar optimization of the design would be required, including considerations of phase velocity and other relevant factors, but the underlying concept remains the same.

Fig.~\ref{fig:energy injection phase sweep}(b) presents the final kinetic electron energy as a function of injection phase, assuming an initial electron energy of 430~keV with the other DLW and THz pulse parameters listed in Table~\ref{tab:DLW parameters} and Table~\ref{tab:THz pulse parameters}.

For this particular design, the optimum injection phase is approximately -14 degrees. Fig.~\ref{fig:energy injection phase sweep} also has relevance for simulating an electron bunch, as electrons within a bunch are injected with varying phases. From Fig.~\ref{fig:energy injection phase sweep}(b), it can be deduced that for larger electron bunches (longitudinally expanded), the optimal injection phase will shift towards phases near zero. This aspect will be further investigated in the last section.

\section{\label{sec:2.Design parameters}Design parameters}

In order to design the DLW and optimize the THz pulse width, we need to fix certain parameters while varying and optimizing others. In the following subsections, we will explore and identify key parameters.  

\subsection{\label{sec:Choice of design parameters}Choice of design parameters}
\textbf{Frequency}: In the first section, we found that phase and group velocity calculations can be extended to different frequencies using a scale factor, provided that the DLW is not excessively lossy. However, due to the non-linear nature of the acceleration process, we cannot simply extend the simulation results in the previous sections to different frequencies. Therefore, we will perform the simulations for some frequency range to understand scaling. 

\textbf{Vacuum radius}: Determining the optimal vacuum radius presents a challenge. On the one hand, a smaller radius results in a higher electric field inside the DLW (see Fig.~\ref{fig:Ez}), but it also leads to a decrease in the acceptable area and electron bunch charge. In addition, the difficulty in fabrication may increase for smaller radii as well as tolerances. Therefore, a compromise must be reached in selecting a vacuum radius that aligns with the designer's and fabrication requirements.

\textbf{Dielectric refractive index}: In this manuscript, we will consider a dielectric refractive index of 1.95 (fused Silica), which remains relatively constant over a wide range of frequencies below 500~GHz \cite{bagdad1968far,tsuzuki2015influence,koike1989optical,naftaly2021terahertz,afsar1984millimeter,dutta1986complex}. It is noted that materials with lower refractive index may require a thicker dielectric to achieve the same phase velocity (lower electric field), while those with higher refractive index may result in a lower group velocity, which is not preferable for a limited duration of the THz pulse. The optimal refractive index may vary depending on the specific parameters.
 
\textbf{Metal conductivity}: We discussed the power loss within the metal and dielectric layer which is not negligible. Therefore, we will take this loss into account in the simulations, resulting in a decreased final kinetic energy compared to the lossless structure. 

\subsection{\label{sec:Varying initial parameters}Varying initial parameters}
\textbf{Initial kinetic electron energy} and \textbf{THz pulse energy}: In order to provide guidelines applicable to a wide range of situations, it is necessary to vary specific initial parameters during the design process. This involves considering the impact of the initial kinetic electron energy and THz pulse energy. 

\subsection{\label{sec:Parameters to be optimized}Parameters to be optimized}

The parameters that need to be optimized include the dielectric thickness, electron injection phase, DLW length, and THz pulse width. It is evident that all these parameters are interconnected and should be optimized collectively. For the following simulations in this section, we just perform a sensitivity analysis for each parameter around its optimum point to study the impact of each one on the final electron energy. All other parameters have been optimized and are listed in Table~\ref{tab:Optimized Parameters}.

\textbf{Electron injection phase}: As explained in the linear acceleration section, the electron must be injected at the leading part of the negative half-cycle to achieve optimal acceleration. It is crucial to optimize the injection phase of the electron to ensure that it remains within the half cycle for as long as possible, experiencing the maximum acceleration while traveling inside the tube. Thus, the injection phase of the electron is optimized in the following simulations to maximize final kinetic electron energy. 

\textbf{Pulse width}: in case of exciting the DLW with a very long THz pulse, the electron falls into the decelerating phase. Conversely, if the THz is too short then the electron outpaces the pulse quickly and cannot be efficiently accelerated. Thus, there exists an optimal pulse width that maximizes acceleration (neglecting DLW dispersion).

\begin{figure}
	\centering
	\includegraphics[width=0.9\linewidth]{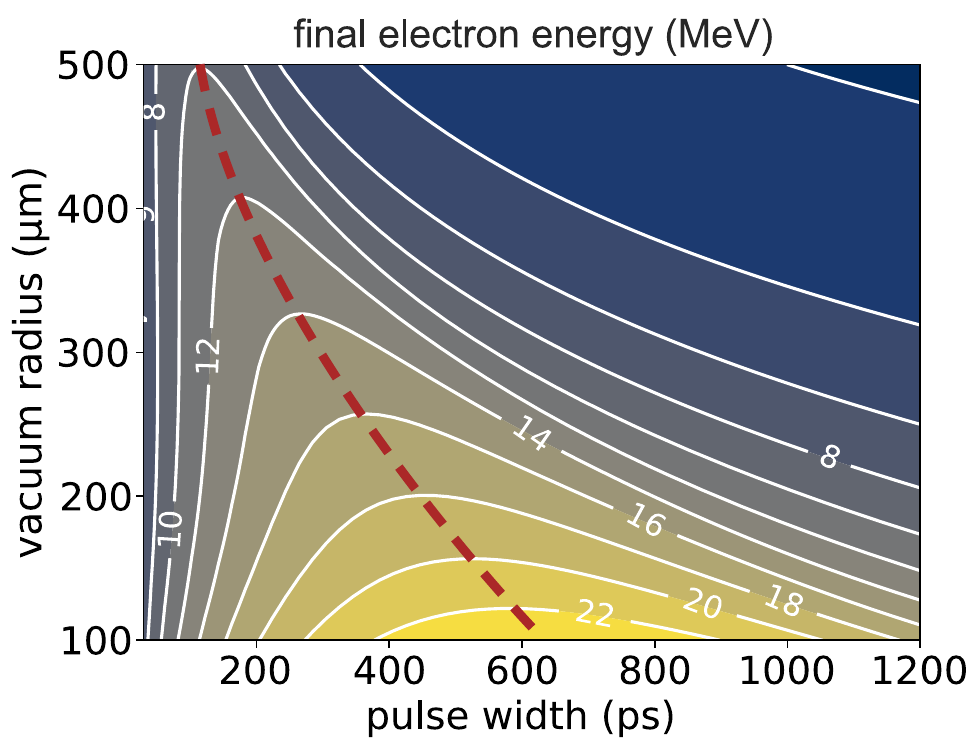}
	\caption{Maximum achievable kinetic electron energy for various vacuum radii and pulse widths, given a 23~mJ THz pulse and a 430~keV initial kinetic electron energy.}
	\label{fig:Optimum_PWO}
\end{figure}

Fig.~\ref{fig:Optimum_PWO} shows the maximum output kinetic energy as a function of vacuum radius and pulse width. Note that the dielectric thickness and the electron injection phase have been optimized for each vacuum radius and pulse width. The other parameters are listed in Table~\ref{tab:DLW parameters} and Table~\ref{tab:THz pulse parameters}.

\textbf{DLW length}: As seen in Fig.~\ref{fig:Optimum_PWO}, there is an optimum pulse width for each vacuum radius to achieve the maximum final electron kinetic energy. The length of the DLW must be chosen in a way that the electrons lead the THz pulse where the DLW ends.
 
\textbf{Dielectric thickness}: By varying the DLW dimensions for a given dielectric, the phase and group velocity are controlled, and there exist optimum dimensions that yield maximum acceleration.

\begin{figure}
	\centering
	\includegraphics[width=0.9\linewidth]{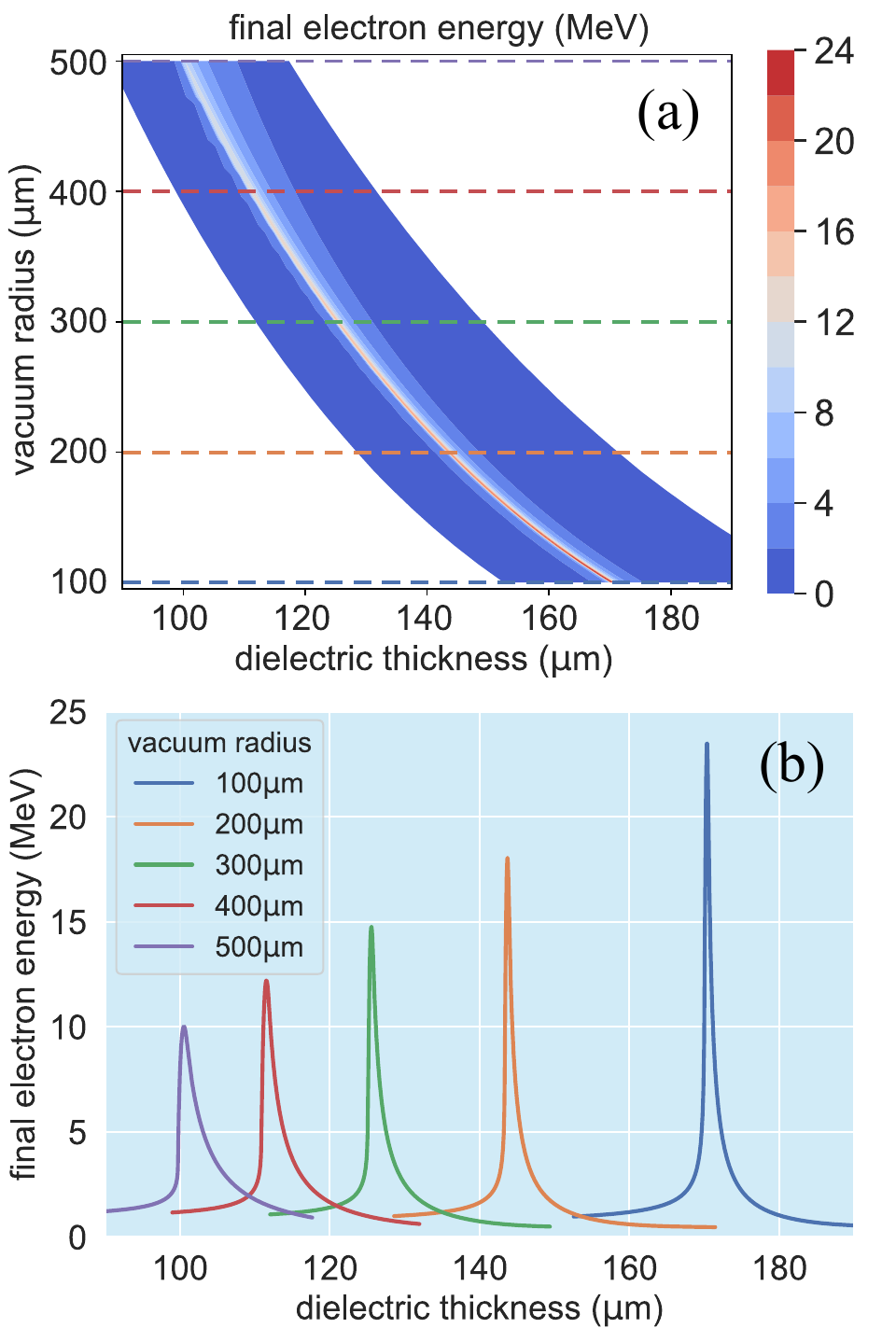}
	\caption{Maximum achievable electron kinetic energy as a function of vacuum radius and dielectric thickness for parameters listed in Table~\ref{tab:DLW parameters} and Table~\ref{tab:THz pulse parameters} (electron injection phase and THz pulse duration are optimized for each vacuum radius and dielectric thickness).}
	\label{fig:VR_DTh_Ek}
\end{figure}

Fig.~\ref{fig:VR_DTh_Ek} illustrates the maximum achievable kinetic energy as a function of the vacuum radius and dielectric thickness for parameters listed in Table~\ref{tab:Optimized Parameters}. Note that the THz pulse width has been optimized by considering the results of Fig.~\ref{fig:Optimum_PWO} for different vacuum radii. The figure shows that reducing the vacuum radius leads to a higher electric field and, subsequently, a higher final electron energy. It's worth noting that there is not an optimum value for the vacuum radius. However, employing a smaller aperture restricts the bunch size and bunch charge. Achieving a balance between these two parameters is beyond the scope of this paper.

\begin{figure*}
	\centering
	\includegraphics[width=1\linewidth]{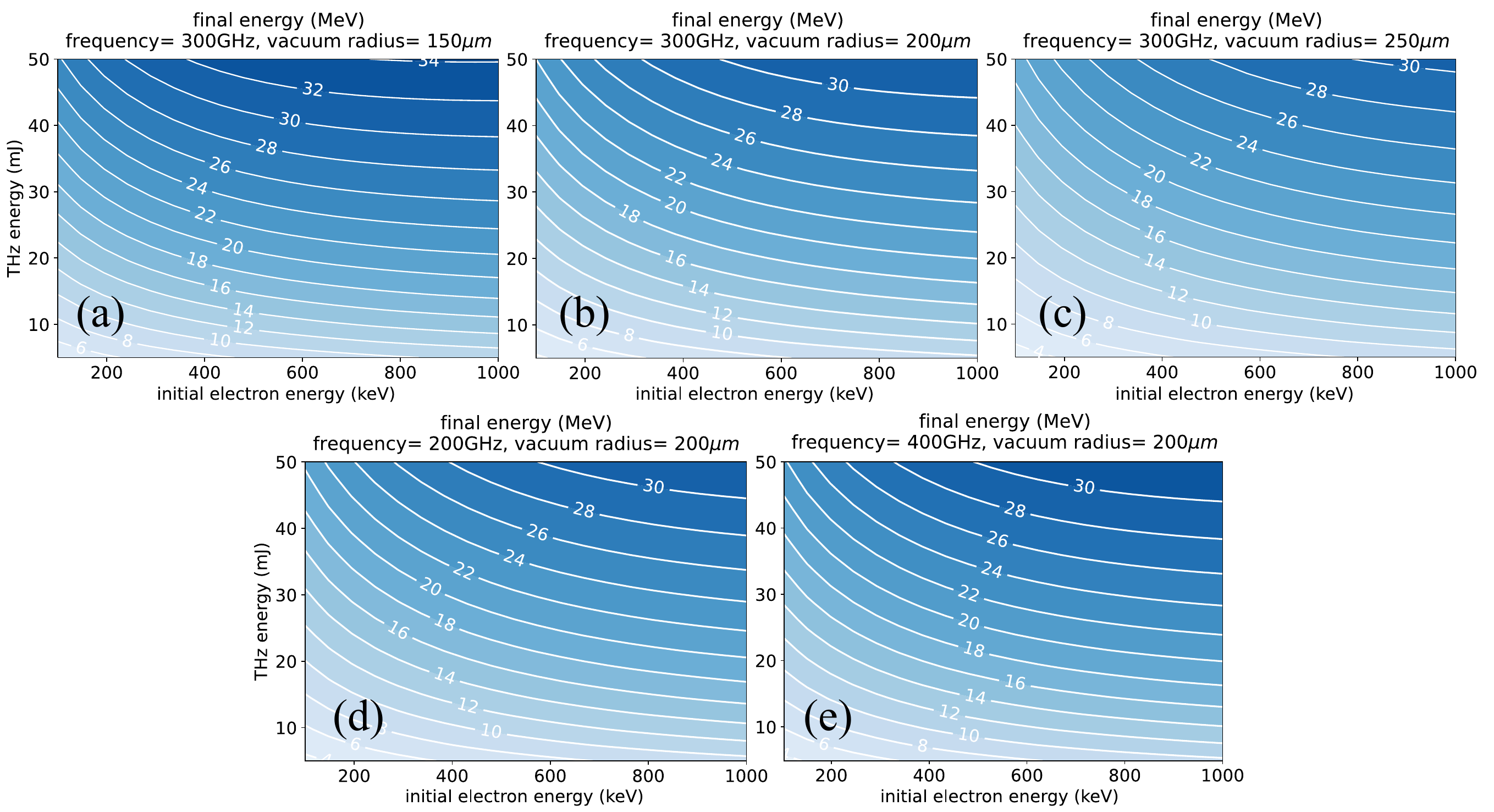}
	\caption{Maximum achievable kinetic energy for different THz pulse energies and initial kinetic electron energies for different vacuum radii of (a) 150~µm (b) 200~µm and (c) 250~µm at 300~GHz and for (d) 200~GHz and (e) 400~GHz THz pulse for the DLW with a vacuum radius of 200~µm assuming always optimum injection phase THz pulse width and DLW length, and dielectric thickness.}
	\label{fig:Ek_PWO}
\end{figure*}

Once all the parameters have been optimized, by sweeping the varying parameters, we can generate guideline figures that enable designers to identify the optimum parameters for a cylindrical DLW LINAC. Fig.~\ref{fig:Ek_PWO} illustrates the maximum achievable kinetic energy for a single electron on the axis of the DLW for parameters listed in Table~\ref{tab:Optimized Parameters}, considering three different vacuum radii and three different frequencies.

\begin{figure*}
	\centering
	\includegraphics[width=1\linewidth]{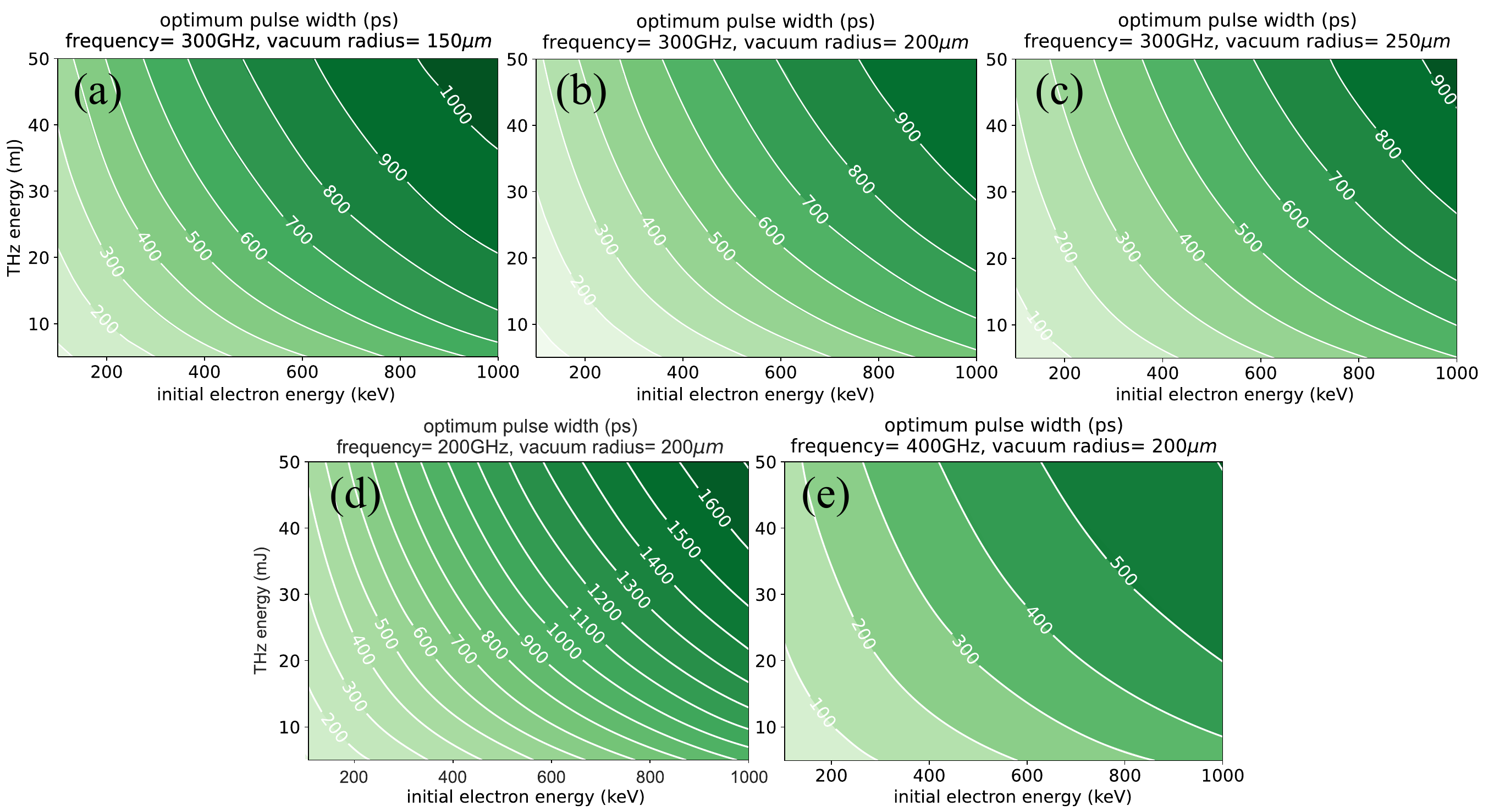}
	\caption{Optimum THz pulse width for different THz pulse energy and initial kinetic electron energy for different vacuum radii of (a) 150~µm (b) 200~µm and (c) 250~µm at 300~GHz and for (d) 200~GHz and (e) 400~GHz THz pulse for the DLW with a vacuum radius of 200~µm assuming always optimum injection phase and dielectric thickness.}
	\label{fig:PWO}
\end{figure*}

These figures are optimized for a single electron. Therefore, for a bunch of electrons, the average kinetic energy might be slightly lower. Fig.~\ref{fig:PWO} displays the optimal width for the THz pulse. It can be observed that a higher energy electron and a higher THz pulse energy lead to a longer optimal pulse width. This is because with less phase slippage, electrons can travel further with the accelerating half-cycle. 

\begin{figure*}
	\centering
	\includegraphics[width=1\linewidth]{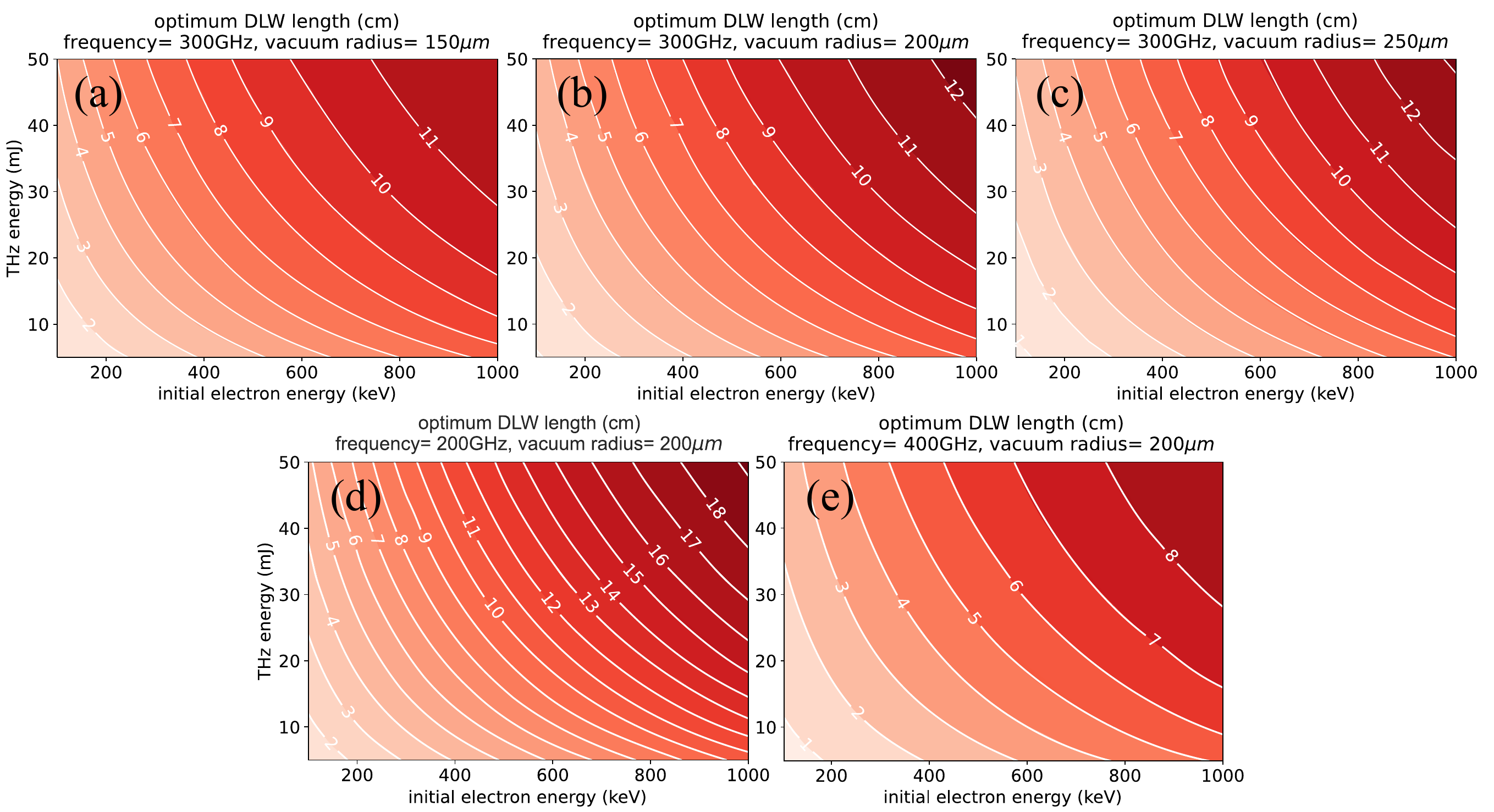}
	\caption{Optimum DLW length for different THz pulse energy and initial kinetic electron energy for different vacuum radii of (a) 150~µm (b) 200~µm and (c) 250~µm at 300~GHz and for (d) 200~GHz and (e) 400~GHz THz pulse for the DLW with a vacuum radius of 200~µm assuming always optimum injection phase and dielectric thickness.}
	\label{fig:Length}
\end{figure*}

Fig.~\ref{fig:Length} shows the optimum DLW length. We can calculate this length based on the optimum pulse width and the group velocity of the THz pulse. 

\begin{figure*}
	\centering
	\includegraphics[width=1\linewidth]{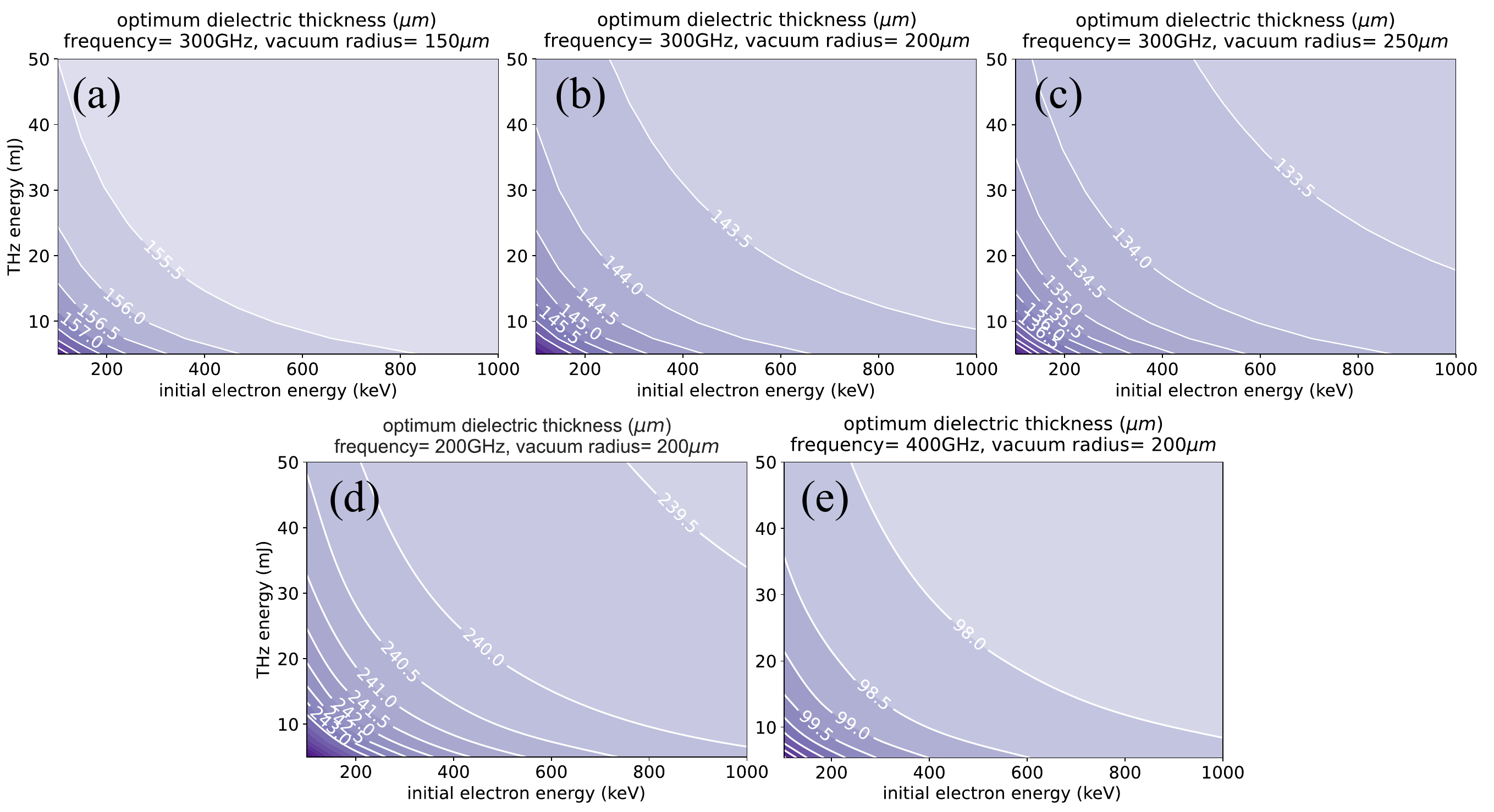}
	\caption{Optimum dielectric thickness for different THz pulse energy and initial kinetic electron energy for different vacuum radii of (a) 150~µm (b) 200~µm and (c) 250~µm at 300 GHz and for (d) 200~GHz and (e) 400~GHz THz pulse for the DLW with a vacuum radius of 200~µm assuming always optimum injection phase, THz pulse width and DLW length.}
	\label{fig:DTh}
\end{figure*}

The phase velocity within the DLW can be controlled by adjusting the dielectric thickness. As depicted in Fig.~\ref{fig:vph and alpha}(a), varying the dielectric thickness results in significant changes in the phase velocity. Consequently, there exists an optimal thickness for each specific combination of THz energy and electron energy. The corresponding optimum values for the dielectric thickness are presented in Fig.~\ref{fig:DTh}.

By utilizing the guideline figures and optimizing the DLW parameters based on initial conditions, designers can customize the DLW LINAC to meet specific requirements. While the simulations presented in this paper are for a single electron on axis of the DLW, it is important to note that the results may vary slightly when simulating an electron bunch. Nevertheless, the designed values obtained from this study serve as a valuable reference and should provide a good starting point for designing and simulating an electron bunch. The optimized parameters offer valuable insights into achieving enhanced performance for THz-driven particle acceleration, making it easier to tailor the DLW LINAC for different scenarios and applications. 

\section{\label{sec:3.Bunch simulation}Bunch simulation}

Up to this point, our focus has been on simulating and optimizing the DLW and THz pulse parameters for a single electron positioned along the DLW axis. In this section, we shift our attention to simulating an electron bunch on the DLW axis to investigate the longitudinal focusing and defocusing effects of a DLW LINAC. Longitudinal focusing is a critical factor in LINACs, as post-focusing is generally unfeasible. In contrast, transverse focusing can be accomplished using DC fields, such as quadrupole magnets. 

When the phase velocity of the THz pulse is set nearly to the speed of light, the longitudinal electric field remains uniform across the aperture, as depicted in Fig.~\ref{fig:Fild distribution}. Furthermore, the magnetic fields influence on longitudinal beam parameters is minimal if the radial electron momentum is negligible. Thus, longitudinal focusing is primarily a result of the electric field gradient, making the simulation of a one-dimensional electron bunch on the DLW axis sufficient for studying axial beam parameters.

As depicted in Fig.~\ref{fig:Fild distribution}(a), due to the alternating positive and negative electric field gradients along the DLW axis, electrons within the phase range of $(2p-1)\frac{\pi}{2}<\phi<2p\frac{\pi}{2}$ experience a focusing field, while those within the range of $2p\frac{\pi}{2}<\phi<(2p+1)\frac{\pi}{2}$ encounter a defocusing field. For a DLW designed for non-relativistic electrons, the phase slippage of electrons relative to the accelerating half-cycle  implies that electrons may experience both focusing and defocusing field during acceleration, as shown in Fig.~\ref{fig:E_z_Phi_z_0_6}(b)). 

One method for controlling the axial beam size involves altering the phase velocity of the THz pulse by varying the THz pulse frequency. Deviating from the optimum design will, naturally, reduce the final kinetic energy. However, keeping the electron bunch within the negative or positive field gradient (rather than at the peak field with zero gradient) for a longer duration could be used for manipulating the longitudinal focus point.

\begin{figure*}
	\centering
	\includegraphics[width=0.9\linewidth]{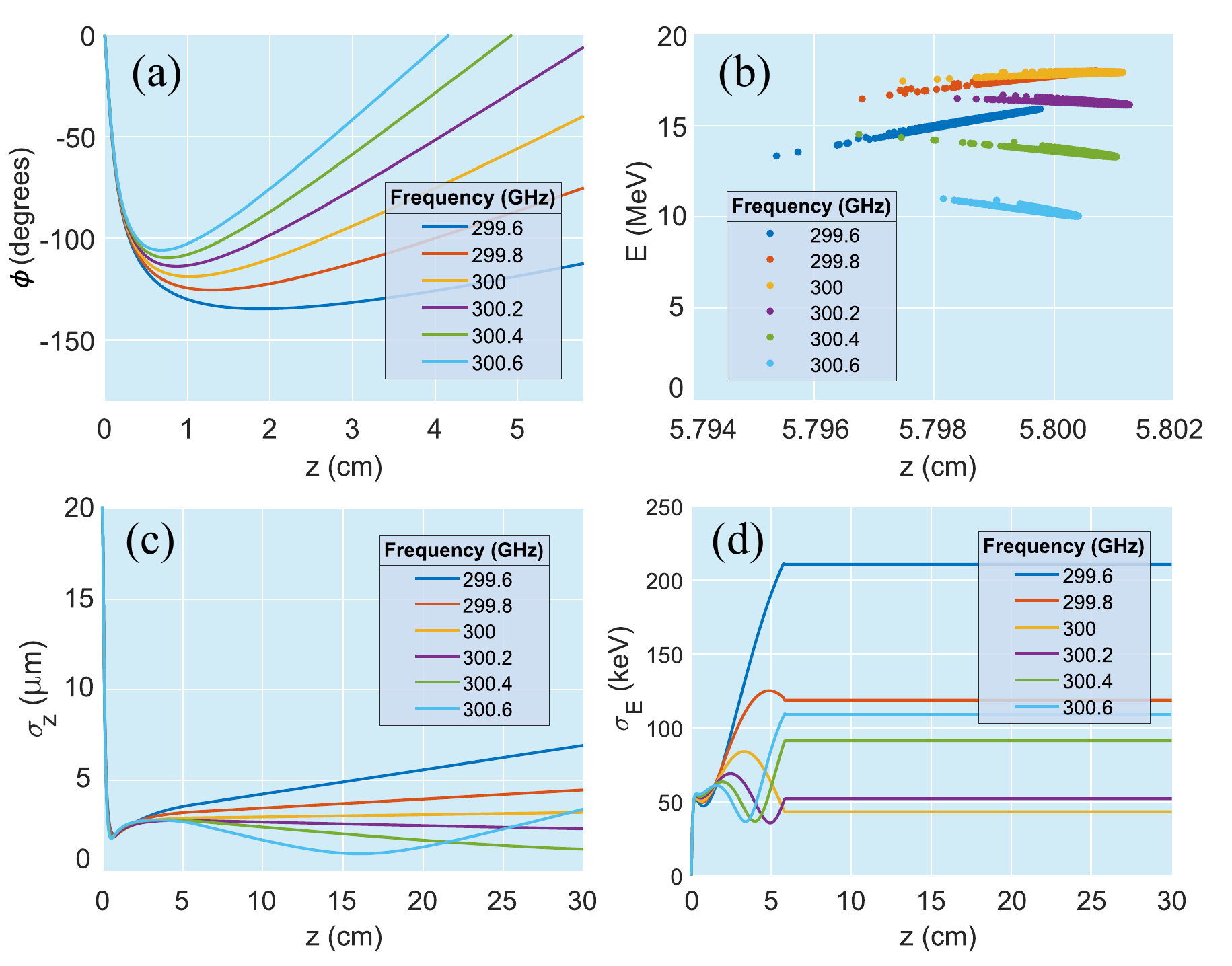}
	\caption{Electron beam parameters for optimized THz pulse and DLW dimension for different THz pulse frequencies. (a) average phase of the electrons (b) electron bunch energy at the end of the LINAC (c) longitudinal beam size (d) energy spread of the electron beam during acceleration.}
	\label{fig:bunch}
\end{figure*}

For our subsequent simulations, we consider an axial Gaussian distribution for the electron bunch with an rms beam size of 20~µm. Fig.~\ref{fig:bunch} displays the bunch parameters, specifically the longitudinal bunch size as a function of distance, for different frequencies. The DLW and THz pulse parameters are kept the same as detailed in Table~\ref{tab:DLW parameters} and Table~\ref{tab:THz pulse parameters}.

Unlike the injection phase optimization for a single electron explained in Fig.~\ref{fig:energy injection phase sweep}(b), for the bunch, we assume a zero-injection phase to maintain all electrons within the accelerating half-cycle.

It is essential to note that since our study primarily focuses on the electromagnetic fields impact on the electron bunch and the bunch charge is generally quite low (below 1~pC), we disregard the space charge effect in our simulations.
All parameters related to the DLW and THz pulse, and the electron bunch are listed in Table~\ref{tab:Optimized Parameters}. 

\begin{table}[ht]
	\caption{\label{tab:Optimized Parameters}Parameters of the optimized DLW and THz pulse}
	\begin{ruledtabular}
	\begin{tabular}{ll}
		\textbf{Parameter} & \textbf{Value} \\
		\midrule
		\multicolumn{2}{l}{\textbf{DLW dimensions}} \\
		\qquad Vacuum radius & 200 µm \\
		\qquad Dielectric thickness & 143.1 µm \\
		\qquad DLW length & 5.8 cm \\
		\multicolumn{2}{l}{\textbf{Material properties}} \\
		\qquad Conductivity of metal & $5.96 \times 10^7$ S/m \\
		\qquad Dielectric refractive index & 1.95 \\
		\qquad Dielectric attenuation constant &\\ 
		\qquad\qquad at 200~GHz & 6 Np/m \\
		\qquad\qquad at 300~GHz & 10 Np/m \\
		\qquad\qquad at 400~GHz & 18 Np/m \\
		\multicolumn{2}{l}{\textbf{THz pulse characteristics}} \\
		\qquad THz pulse frequency & 300 GHz \\
		\qquad THz pulse energy & 23 mJ \\
		\qquad THz pulse duration & 467 ps \\
		\multicolumn{2}{l}{\textbf{Electron bunch characteristics}} \\
		\qquad Average injection phase & 0 degree \\
		\qquad Longitudinal rms bunch size & 20 µm \\
	\end{tabular}
	\end{ruledtabular}
\end{table}

Fig.~\ref{fig:bunch} shows the electron bunch parameters during acceleration for optimized THz pulse and DLW dimension listed in Table~\ref{tab:Optimized Parameters} for different THz pulse frequencies around 300~GHz.

Fig.~\ref{fig:bunch}(c) illustrates a substantial reduction in the initial electron beam size, primarily in the DLW's early section, due to the low momentum of the electrons. For the optimal frequency (300~GHz), the output beam is approximately collimated. An increase in frequency results in a reduction of the phase velocity. Consequently, the electron bunch remains in the positive field gradient for a longer duration, leading to beam convergence. This can be inferred from the electron beam energy at the exit of the DLW (Fig.~\ref{fig:bunch}(b)) where electrons at the front of the bunch exhibit lower energy and velocity (negative slope). In certain applications, depending on the components downstream of the LINAC, a more tightly converged electron beam may be required. The simulations conducted indicate that achieving this can be accomplished by adjusting the THz frequency. However, it's important to note that such adjustments will result in a reduction in the final energy of the electron beam.
It's worth noting that lower frequencies have the potential to maintain the bunch quality due to the gradual slope of the wavelength. Conversely, higher frequencies enable higher field gradients. 

\FloatBarrier
\section{\label{sec:Conclusion}Conclusion}
We presented a comprehensive analytical/numerical guide for optimizing cylindrical dielectric loaded waveguide (DLW) parameters in the context of terahertz (THz) driven linear accelerators (LINACs). By carefully adjusting the dimensions of the dielectric and THz pulse properties, it is possible to control the phase and group velocities within the DLW, and making an effective interaction between the electron and the THz pulse. Thus, optimum electron acceleration can be achieved. The research has identified critical parameters that need to be optimized, such as dielectric thickness, DLW length, THz pulse duration, etc., which play significant roles in maximizing the final kinetic energy of accelerated electrons. What is also remarkable is the relatively weak sensitivity of the resulting electron energy as a function of the frequency used, given a certain THz pulse energy (see Figs.~\ref{fig:Ek_PWO}-\ref{fig:DTh}). The reason is that while, for high THz frequencies, the DLW cross-section becomes smaller, resulting in a higher field at the same intensity and thus allowing for a higher gradient acceleration, the dephasing length and attenuation constant simultaneously becomes shorter and higher. Consequently, the length of the accelerating section powered is reduced. Therefore, at fixed THz energy, the final electron energy achievable increases only slowly with increasing frequency. The presented guideline figures serve as valuable tools for designers to tailor the accelerator to specific requirements, offering insights into the performance enhancement of THz-driven particle acceleration. Although the simulations are conducted for a single electron on the axis of a DLW and longitudinally spread electron bunches, the obtained optimized values can be considered as a reliable starting point for designing and simulating electron bunches. 

\begin{acknowledgments}
The authors acknowledge support by DESY, a Center of the Helmholtz Association, the European Research Council under the European Union’s Seventh Framework Programme (FP7/2007-2013) through the Synergy Grant AXSIS (609920), the Cluster of Excellence 'Advanced Imaging of Matter' of the Deutsche Forschungsgemeinschaft (DFG) - EXC 2056 - project ID 390715994 and Project 655350 of the Deutsche Forschungsgemeinschaft (DFG) and the accelerator on a chip program (ACHIP) funded by the Gordon and Betty Moore foundation (GBMF4744).
\end{acknowledgments}

\bibliography{apssamp}

\end{document}